\documentclass[prl,aps,twocolumn,showpacs]{revtex4}

\usepackage{ifpdf}
\ifpdf
\usepackage{subfigure}
\usepackage[pdftex]{graphicx}
\else
\usepackage{graphicx}
\fi

\usepackage{setspace}
\usepackage{psfrag}
\usepackage{epstopdf}
\usepackage{graphicx}
\DeclareGraphicsExtensions{.eps}

\usepackage{amsmath,graphicx,epsfig,bm,color}
\usepackage{amssymb,amsfonts}
\usepackage{rotating}
\usepackage{graphics}
\usepackage{setspace}

\begin{document}
%\numberwithin{equation}{section}
%\def\pd#1#2{\frac{\partial #1}{\partial #2}}

\title{\bf A classification of the ground states and topological defects in a rotating two-component Bose-Einstein condensate}
\author {Peter Mason${}^{1,2}$ and Amandine Aftalion${}^3$}
\affiliation {${}^1$Laboratoire de Physique Statistique, \'Ecole Normale Sup\'erieure, UPMC Paris 06, Universit\'e Paris Diderot, CNRS, 24 rue Lhomond, 75005 Paris, France\\
${}^2$Institut Jean Le Rond D'Alembert, UMR 7190 CNRS-UPMC, 4 place Jussieu, 75005 Paris, France\\
${}^3$CNRS \& Universit\'e Versailles-Saint-Quentin-en-Yvelines,
Laboratoire de Math\'ematiques de Versailles, CNRS UMR 8100, 45
avenue des \'Etats-Unis, 78035 Versailles C\'edex, France}
\date{\today}

\begin{abstract}
%This paper considers two-dimensional two-component condensates with distinct intracomponent coupling strengths.
 We classify the ground states and topological defects of a rotating
   two-component condensate when varying several parameters: the intracomponent
  coupling strengths, the intercomponent coupling strength and the particle numbers.
  %and rotation is taken within the physically permissible range.
%($0\rightarrow\omega$, for trap frequency $\omega$).
 No restriction is placed on the masses or trapping
frequencies of the individual components. We present numerical phase diagrams
 which show the boundaries between the regions of
coexistence, spatial separation and symmetry breaking. Defects such
as triangular coreless vortex lattices, square coreless vortex
lattices and giant skyrmions are classified. Various aspects of the
phase diagrams are analytically justified thanks to a non-linear $\sigma$-model
that describes the condensate in terms of the total density and
 a pseudo-spin representation.
%The total density profile of the condensate in the Thomas-Fermi limit and in the
%absence of topological defects can be determined purely from this non-linear $\sigma$-model.
\end{abstract}

%\pacs{??}
\maketitle

\section*{I. Introduction}

Bose Einstein condensates (BECs) provide
 an excellent environment to study experimentally and theoretically
 a rich variety of macroscopic
quantum phenomena. In a rotating single component condensate,
topological defects to the order parameter often manifest themselves as vortices that
correspond to a zero of the order parameter with a circulation of
the phase. When they get numerous, these vortices  arrange
themselves on a triangular lattice.
In fact, vortices were first
observed in two-component BEC's \cite{matthews}. Since then two-component BECs and the topological excitations within have been experimentally realised in a number of configurations: a single isotope that is in two different
hyperfine spin states \cite{hall,ander,myatt,delannoy,madd,matthews,sch}, two different isotopes of the same
atom \cite{papp} or isotopes of two different atoms \cite{modugno,thall,ferrari}.
%Two-component Bose-Einstein condensates  can represent various types of configuration: A two-component BEC (equivalently a {\sl two-species}, {\sl binary}, {\sl dual-species} or {\sl Bose-Bose} condensate) has been realised in a number of experimental papers in each of these different configurations \cite{thall,stamper,papp,myatt,modugno,miesner,hall,ferrari,delannoy,lew,madd,ander}.

When a two-component
condensate is under rotation, topological defects of both order
parameters are created which lead to more exotic defects such as
singly or multiply quantised skyrmions
 \cite{wh,sch,mertes,cooper,savage,ruo}. Analogy with the Skyrme model
  from particle physics is often invoked to represent the defects \cite{manton,cho}. The
singly quantised skyrmions contain a vortex in one component which
has the effect of creating a corresponding density peak in the other
component. These singly quantised skyrmions are often referred to as
coreless vortices. Once numerous, these vortices and peaks arrange
themselves in (coreless vortex) lattices, that can be either
triangular or square. Other defects such as vortex sheets or giant
skyrmions can also be observed \cite{ktu,ktu1,ywzf}.  The aim of this
paper is to classify the type of defects according to the different
parameters of the problem restricting ourselves to two-dimensions. While this
paper will only concern itself with magnetically trapped
two-component BEC's, there is active research into
%three-component or
spinor condensates (for a recent review see
\cite{uk}).

   In the mean-field regime, the two-component Bose-Einstein condensate
at zero temperature is described in terms of two  wave functions (order parameters),
$\Psi_1$ and $\Psi_2$, respectively representing component-1 and
component-2.
 The two-component condensate is placed into rotation about the $z$-axis with $\bm{\bar{\Omega}}=\bar{\Omega}\bm{\hat{z}}$ where $\bar{\Omega}$ is the rotation frequency assumed to apply equally to both components. The Gross-Pitaevskii (GP) energy functional of the rotating two-component, two-dimensional BEC is then given by
\begin{equation}
\begin{split}
\label{enfn}
E[\Psi_1,\Psi_2]=\int \sum_{k=1,2}\Bigg(\frac{\hbar^2}{2m_k}|\nabla\Psi_k|^2+V_k(\bm{r})|\Psi_k|^2\\
-\hbar\bar{\Omega}\Psi^*_k L_z \Psi_k +\frac{U_k}{2}|\Psi_k|^4\Bigg)+U_{12}|\Psi_1|^2|\Psi_2|^2\quad d^2r
\end{split}
\end{equation}
where $r^2=x^2+y^2$, and  $V_k(\bm{r})=m_k\omega_k^2r^2/2$ are the
harmonic trapping potentials with trapping frequencies $\omega_k$,
centered at the origin. Here $m_1$ and $m_2$ are the masses of the
bosons in component-1 and component-2 respectively, and $\hbar$ is
Planck's constant. The angular momentum is in the $z$-axis and is
defined as $L_z=i[\bm{\hat{z}}\cdot(\bm{r}\times\bm{p})]$ for linear
momentum $\bm{p}$.
% In general, the problem is three dimensional but if the characteristic energy of the harmonic oscillator in the $z$ direction  is large with respect to the  other energy scales, then the  atoms
% can be assumed to occupy the ground state of the harmonic oscillator in this direction, so that the geometry is restricted to two dimensions.
    The energy functional (\ref{enfn})
 contains three interaction constants:
 $U_k$ representing the internal interactions in  component $k$, and $U_{12}$
  representing the interactions between the two components.

    The time-dependent coupled Gross-Pitaevskii (GP) equations
    are obtained from a variational procedure, $i\hbar\partial\Psi_k/\partial t=\delta E/\delta\Psi^*_k$,
     on the energy functional (\ref{enfn}).
%\begin{eqnarray}
%i\hbar\frac{\partial\Psi_k}{\partial
%t}&=&-\frac{\hbar^2}{2m_k}\nabla^2\Psi_k+\frac{1}{2}m_k\omega_k^2r^2\Psi_k+U_k|\Psi_k|^2\Psi_k\nonumber\\&\quad&+U_{12}|\Psi_{3-k}|^2\Psi_k-\hbar\bar{\Omega}
%L_z\Psi_k. \label{gp1}
%\end{eqnarray}
Let $\tilde{\omega}=(\omega_1+\omega_2)/2$ be the average of the
trapping frequencies of the two components and introduce the reduced mass $m_{12}$  such that
$m_{12}^{-1}=m_1^{-1}+m_2^{-1}$. The GP energy and the coupled GP
equations  can be non-dimensionalised by choosing
$\tilde{\omega}^{-1}$, $\hbar\tilde{\omega}$ and
$r_0=\sqrt{\hbar/(2m_{12}\tilde{\omega})}$ as units of time, energy
and length respectively.  On defining the non-dimensional
intracomponent coupling parameters $g_k=2U_km_{12}/\hbar^2$ and the
intercomponent coupling parameter $g_{12}=2U_{12}m_{12}/\hbar^2$
($\equiv g_{21}$), the dimensionless coupled GP equations read
    \begin{equation}
    \begin{split}
\label{gp11}
%i\frac{\partial\psi_k}{\partial t}&=&-\frac{m_{12}}{m_k}\nabla^2\psi_k+\frac{m_k}{4m_{12}}\frac{\omega_k^2}{\tilde{\omega}^2}r^2\psi_k+g_k|\psi_k|^2\psi_k\nonumber\\&\quad&+g_{12}|\psi_{3-k}|^2\psi_k-\Omega L_z\psi_k,
i\frac{\partial\psi_k}{\partial t}=&-\frac{m_{12}}{m_k}\left(\nabla-i\bm{A}_k\right)^2\psi_k+\frac{m_k}{4m_{12}}\left(\frac{\omega_k^2}{\tilde{\omega}^2}-\Omega^2\right)r^2\psi_k\\
&\quad+g_k|\psi_k|^2\psi_k+g_{12}|\psi_{3-k}|^2\psi_k,
\end{split}
\end{equation}
for $\Omega=\bar{\Omega}/\tilde{\omega}$ and where
\begin{equation}
    \bm{A}_k=\frac{1}{2}\frac{m_k}{m_{12}}\bm{\Omega}\times\bm{r},
\end{equation}
for $\bm{\Omega}=(0,0,\Omega)$ and $\bm{r}=(x,y,0)$.  The ground
state of the energy (\ref{enfn}) or the GP equations (\ref{gp11}) is
determined by preserving the normalisation condition which in this
paper is either taken to be
\begin{equation}
\label{norm} \int |\psi_k|^2\quad d^2r=N_k,
\end{equation}
where $N_k$ is the total particle number of the $k$th-component, or
\begin{equation}
\int \left (|\psi_1|^2+|\psi_2|^2\right )\quad d^2r=N_1+N_2.
\label{norm12}
\end{equation}

The following sections will only consider repulsive interactions so
that $g_1$, $g_2$ and $g_{12}$ are always non-negative.
%We will see that the sign of $ g_2- \Lambda g_1$ will be important: this will be clarified in Sect. IV, but we here define $\Lambda$ as
%\begin{equation}
%   \Lambda=\left(\frac{V_2^{\text{eff}}(1)}{V_1^{\text{eff}}(1)}\right)^2=\frac{(4-(1+\xi)^2\Omega^2)^2}{\eta^2(4\xi^2-(1+\xi)^2\Omega^2)^2}.
%   \end{equation}
In order to separate the regimes of interest, a non-dimensional
parameter combining these $g_k$ and $g_{12}$ will appear:
%The following non-dimensional parameter is also introduced
%\begin{subequations}
\begin{equation}\label{gamma12}
%\begin{eqnarray}
%    \Gamma_k&=&1-\frac{g_{12}}{g_k},\\
    \Gamma_{12}=1-\frac{g_{12}^2}{g_1g_2}.
\end{equation}
%\end{eqnarray}
%\end{subequations}
Furthermore to simplify matters, we introduce ratio parameters $\eta$ and $\xi$ such that
\begin{equation}m_1=\eta m_2\hbox{  and }\omega_1=\xi\omega_2\hbox{ with
}\{\eta,\xi\}>0.\end{equation} The effective trapping potentials
 for each component coming from equation (\ref{gp11}) are then,
respectively,
\begin{subequations}
\begin{align}
    V_1^{\text{eff}}(r)=&(\eta+1)\left(\frac{\xi^2}{(\xi+1)^2}-\frac{1}{4}\Omega^2\right)r^2,\\
    V_2^{\text{eff}}(r)=&\frac{(\eta+1)}{\eta}\left(\frac{1}{(\xi+1)^2}-\frac{1}{4}\Omega^2\right)r^2.
\end{align}
\end{subequations}
%where $\gamma_1=(\eta+1)/(\xi+1)^2$ and $\gamma_2=\xi^2/\eta$.
It follows that the limiting rotational frequency for each component
is $\Omega_1^{\text{lim}}=2\xi/(\xi+1)$ and
$\Omega_2^{\text{lim}}=2/(\xi+1)$, so that it is necessary to
consider a rotational frequency  such that
$0<\Omega<\Omega^{\text{lim}}=\min\{{\Omega_1^{\text{lim}},\Omega_2^{\text{lim}}}\}$.

Experimentally, it is often the case that the $U_k$, $m_k$,
$\omega_k$ and $N_k$  are of the same order, so that much of the
theoretical and numerical analysis concerning two-component
condensates has assumed equality of these parameters.
 In the case where the
intracomponent coupling strengths are equal, Kasamatsu et al.
\cite{ktu,ktu1} produced a numerical phase diagram in terms of the
rotation and the intercomponent coupling strengths: they found
 phase separation regions with either vortex sheets
or droplet behavior and regions of coexistence of the components
with coreless vortices, arranged in triangular or square lattices.

In this paper, we present a  classification of the ground states and
various types of topological defect when the intracomponent coupling
strengths are distinct. Depending on the magnitude of the various
parameters the components can either coexist, spatially separate
or exhibit symmetry breaking. A richer phenomenology of topological
defects is then found.

Much of the analysis carried out to investigate the ground states and topological defects will use a
%Some analysis will be carried out using a
 a nonlinear sigma model.
 It has been introduced previously in the literature \cite{ktu,ktu3} for $\eta=\xi=1$ but
%here we manage to find a generalized formulation.
we will generalise this to the cases $\eta$ and $\xi$ different from
1. This  involves writing the total density
as
\begin{equation}\label{rhot}
    \rho_T=|\psi_1|^2+\eta|\psi_2|^2.
\end{equation}
  A normalised complex-valued spinor
$\bm{\chi}=[{\chi_1},{\chi_2}]^T$ is introduced from which the
wave functions are decomposed as $\psi_1=\sqrt{\rho_T}\chi_1$,
$\psi_2=\sqrt{\rho_T/\eta}\chi_2$ where $|\chi_1|^2+|\chi_2|^2=1$.
 Then we define the spin density $\bm{S}=\bar{\bm{\chi}}\bm{\sigma}\bm{\chi}$ where
$\bm{\sigma}$ are the Pauli matrices. This gives the components of $\bm{S}$ as
%$\bm{\underline{\underline{\sigma}}}$
\begin{subequations}
    \label{seqs}
\begin{align}
    \label{sx}
    S_x=&\chi^*_1\chi_2+\chi_2^*\chi_1,\\
    \label{sy}
    S_y=&-i(\chi^*_1\chi_2-\chi_2^*\chi_1),\\
    S_z=&|\chi_1|^2-|\chi_2|^2,
    \label{sz}
\end{align}
\end{subequations}
with $|\bm{S}|^2 = 1$ everywhere. We will write the GP energy (\ref{enfn}) in terms of
 $\rho_T$ and $\bm{S}$ that will allow us to derive information on
 the ground state of the system.

The paper is organised as follows. The different regions of the
$\Omega-\Gamma_{12}$ phase diagrams are outlined in Sect. II with a
detailed analysis of the range of ground states and topological
defects. Then the non-linear $\sigma$-model is developed in Sect.
III to analyse the ground states in terms of the total density. Lastly
Sect. IV  takes the second normalisation condition (\ref{norm12}) and presents an example $\Omega-\Gamma_{12}$ phase diagram.

\section*{II. The $\Gamma_{12}-\Omega$ Phase Diagram}

\subsection*{A. Numerical Parameter Sets}

 In this paper,
three different configurations are considered, two of which relate directly to experimental configurations.
 Firstly, we analyze a
$^{87}$Rb-$^{87}$Rb mixture  with one isotope in spin state $|F=2$,
$m_f=1\rangle$ and the other in state $|1$, $-1\rangle$. The complex
relative motions between these two isotope components of rubidium
were considered experimentally by Hall et al \cite{hall}. Here, the
masses are equal ($\eta=1$) and  the transverse trapping potentials
 are equal, $\omega_1=\omega_2$ ($\xi=1$). The scattering lengths for each
component are $a_1=53.35\AA$ and $a_2=56.65\AA$.
  The intracomponent coupling strengths
   $U_k$ for a two-dimensional model are  defined as
%\begin{subequations}\begin{equation}
$U_k={\sqrt{8\pi}\hbar^2a_k}/[{m_ka_{zk}}]$ ($k=1,2$),
%\end{equation}
and the intercomponent coupling strength is defined as
%\begin{equation}
$U_{12}={\sqrt{2\pi}\hbar^2a_{12}}/[{m_{12}\tilde{a}_z}]$,
%\end{equation}
%\end{subequations}
where
 $m_{12}$ is the reduced mass, given by $m_{12}^{-1}=m_1^{-1}+m_2^{-1}$,
$a_k$, $a_{12}$ are the s-wave
(radial) scattering lengths,
$a_{zk}$ is the characteristic size of the condensate in the
  $z$ direction,
  %that is $a_{zk}=\sqrt{\hbar/m_{k}\omega_{zk}}$ with
 %$\omega_{z1}$ and $\omega_{z2}$  the frequencies of the confinement in the $z$ direction,
and $\tilde{a}_z=(a_{z1}+a_{z2})/2$.
 In the experiment \cite{hall}, the computations of approximate values of $a_{z1}$
 and $a_{z2}$ can be made using a Thomas-Fermi approximation in the $z$ direction
  as explained in \cite{PS}, chapter 17. This relies on the assumption
  that $N_k a_k/a_{zk}$ are large. Then, it is reasonable to describe the experiments  through a two-dimensional model.
 For our simulations, we choose $g_1=0.0078$ and
$g_2=0.0083$, which are consistent with the experimental data and we set $N_1=N_2=10^5$.  We denote
this parameter set as `Experimental Set 1' (ES1).

Secondly, we tackle a $^{41}$K-$^{87}$Rb mixture with both isotopes
in spin state $|2$, $2\rangle$, as was considered by Modugno et al
\cite{modugno}. Component-1 is identified with
the $^{41}$K isotope and component-2 with the $^{87}$Rb isotope. In the experiment,
  the masses are different ($\eta\sim 0.48$),
  but since $\eta\xi^2=1$, then $m_1 \omega_1^2= m_2\omega_2^2$
   so that  both components experience the
same trapping potential.
%$\omega_{z2}=16.3\times2\pi$\phantom{ }Hz, $\omega_2=190\times2\pi$\phantom{ }Hz and
%$\{\omega_{z1},\omega_1\}=\xi\{\omega_{z2},\omega_2\}$
Here the scattering lengths for each component are $a_1=31.75\AA$
and $a_2=52.39\AA$ and we  choose intracomponent coupling  strengths
as $g_1=0.0067$ and $g_2=0.0063$. This set is denoted ES2.

 Lastly, in order to make an
analogy with previous theoretical studies, we consider the  set
(ES3) which contains a mixture chosen such that all the parameter
groups are equal, i.e. here $g_1=g_2=0.0078$ and $N_1=N_2=10^5$ with
equality of the $m_k$ and
$\omega_k$ ($\eta=\xi=1$). %Then $\Lambda=1$ and $g_2-\Lambda g_1=0$.

In each case, the features of the ground states will be explained as
$g_{12}$ and $\Omega$ are varied. While the value of the $g_k$ are
nominally fixed, the value of the intercomponent coupling strength,
$g_{12}$, can be altered by Feshbach resonance (see for instance
\cite{tojo,theis,bauer,kempen}), which allows for an
extensive experimental range in $\Gamma_{12}$, given by (\ref{gamma12}) (keeping $\Gamma_{12}\le1$). Simulations
are thus performed on the coupled GP equations (\ref{gp11}) in
imaginary time. In general, it is difficult to find the true minimizing energy state. But the use of various initial data converging to the same final state allows us to state that the true ground state will be of the same pattern as the one we exhibit.

 We would like to note that while we have chosen these particular parameters, we have conducted extensive numerical simulations over a range of different parameters sets and find the sets presented here illustrate well the possible ground states.

%\subsection*{B. Non-rotating Ground State Profiles}
\subsection*{B. Classification of the Regions of the Phase Diagram}

The ground states can be classified according to
\begin{enumerate}
\item the symmetry properties
\item the properties of coexistence of the condensates or spatial
separation.
\end{enumerate}

\begin{figure}
\begin{center}
\includegraphics[scale=0.5]{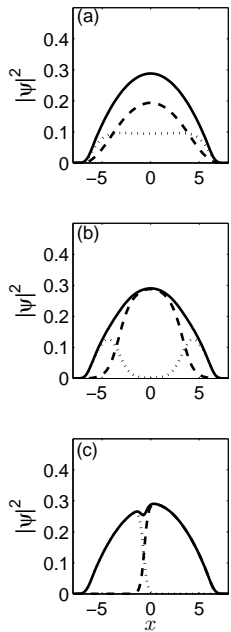}
\end{center}
\caption{Density (divided by $10^4$) plots along
$y=0$ for component-1 (dashed lines), component-2 (dotted lines) and
the total density (solid lines) in which the components (a) coexist
both being disks ($\Gamma_{12}=0.1$), (b) spatially separate to be a
disk and an annulus ($\Gamma_{12}=0$) and (c) have symmetry broken
to be two droplets ($\Gamma_{12}=-0.3$). The parameters are
$g_1=0.0078$, $g_2=0.0083$ and $N_1=N_2=10^5$ with $\eta=\xi=1$ (set ES1). The
angular velocity of rotation is $\Omega=0$. Distance is measured in units of $r_0$ and density in
units of $r_0^{-2}$.} \label{gs}
\end{figure}

When there is no rotation ($\Omega=0$), as illustrated
 in Fig. \ref{gs}, the geometry of the ground state can either be
 \begin{itemize}\item two disks (coexistence of the components and symmetry preserving state)
 \item a disk and an annulus (symmetry preserving with  spatial
 separation of the components), which depends strongly on the fact that the $g_k$, $m_k$, $\omega_k$ are
not equal
 \item droplets (symmetry breaking and spatial separation).
 \end{itemize}

\begin{figure}
\begin{center}
\includegraphics[scale=0.5]{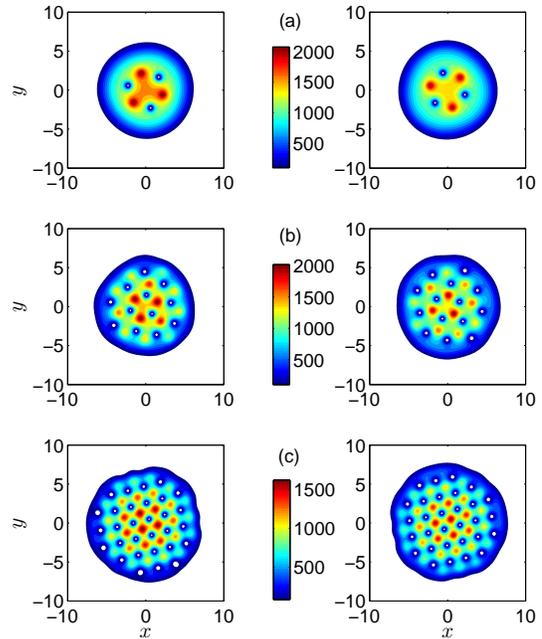}
\end{center}
\caption{(Color online) A series of density plots for
component-1 (left column) and component-2 (right column) in which
the components coexist and are both disks. The parameters are
$g_1=0.0078$, $g_2=0.0083$ and $N_1=N_2=10^5$ and with $\eta=\xi=1$ (set ES1)
and $\Gamma_{12}=0.5$ (which gives $g_{12}=0.0057$). The angular
velocity of rotation is $\Omega$ and it takes the values (a) $0.25$,
(b) $0.5$, and (c) $0.75$. Vortices in one component have a
corresponding density peak in the other component (coreless
vortices). In  (c) the coreless vortex lattice is square. At these
parameters the components are in region 1 of the phase diagram of
Fig. \ref{sch_2}. Distance is measured in units of $r_0$ and density in
units of $r_0^{-2}$.} \label{05}
\end{figure}

% \begin{figure}
% \begin{center}
% \includegraphics[scale=0.5]{m_d.eps}
% \end{center}
% \caption{(Color online) A density plot for component-1 (left column) and component-2 (right column) showing examples of multiple vortex nucleation in component-2. The parameters are $g_1=0.0067$, $g_2=0.0063$ and $N_1=N_2=10^4$ with $\eta\sim0.48$, $\eta\xi^2=1$ (set ES2), $\Gamma_{12}=0.7$ (which gives $g_{12}=0.0036$) with (a) $\Omega=0.5$ and (b) $\Omega=0.75$. At these parameters the components are in region 1 of the phase diagram of Fig. \ref{sch_3}. Distance is measured in units of $r_0$ and density in
% units of $r_0^{-2}$.}
% \label{m_d}
% \end{figure}

Under rotation, the different ground states
%taking into account the symmetry or breaking of symmetry and the type of defects
can be classified according to the parameters  $\Gamma_{12}$ and
$\Omega$. When the condensates are two disks,  or a disk and annulus, defects may break some symmetry of the system as $\Omega$ is increased. But the wave function retains some non trivial rotational symmetry. We will not refer to this as symmetry breaking since the bulk
 condensate keeps some symmetry. We will use the terminology symmetry breaking when the bulk completely breaks the symmetry of the system and is a droplet or has vortex sheets.
 We find that there are four distinct regions, determined by the geometry of the ground state.

{\em Region 1}: Both components are disk shaped, with no spatial separation. Above some critical velocity
$\Omega$, coreless vortices appear:
 a vortex in one component which has the effect of
creating a corresponding density peak in the other component. They
arrange themselves either on a triangular or on a square lattice.
Figure \ref{05} provides a typical example of the density profiles.

\begin{figure}
\begin{center}
\includegraphics[scale=0.5]{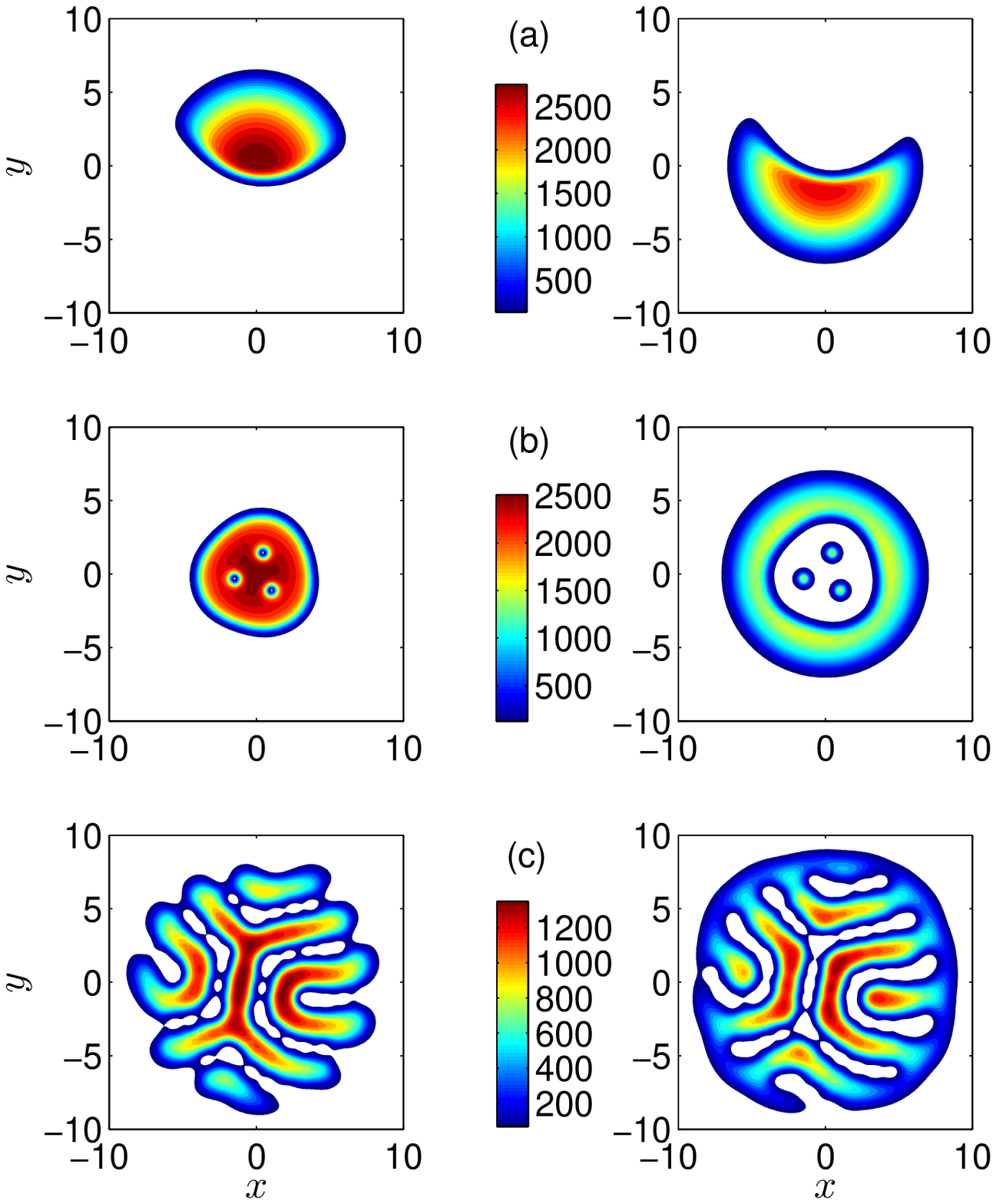}
\end{center}
\caption{(Color online) A series of density plots for
component-1 (left column) and component-2 (right column) in which
the components have spatially separated. The parameters are
$g_1=0.0078$, $g_2=0.0083$ and $N_1=N_2=10^5$ and with $\eta=\xi=1$ (set ES1)
 and $\Gamma_{12}=-0.3$ (which gives
$g_{12}=0.0092$). The angular velocity of rotation is $\Omega$ and
it takes the values (a) $0.1$, (b) $0.5$, and (c) $0.9$. In (a) the
components are rotating droplets (region 4 of Fig. \ref{sch_2}), in
(b) the components have spatially separated (but keep some symmetry) and have isolated density peaks (region 3 of Fig.
\ref{sch_2}) and in (c) there are vortex sheets present (region 2 of
Fig. \ref{sch_2}). Distance is measured in units of $r_0$ and density in
units of $r_0^{-2}$.} \label{03neg}
\end{figure}

{\it Region 2}: Vortex sheets. Under the effect of strong rotation,
the components spatially separate and completely break the symmetry of the system. Nevertheless, they are approximately
 disk shaped with similar radii. Many vortices are nucleated that
arrange themselves into rows  that can have various patterns: either
they can be striped or bent and are often disconnected from each
other. A vortex sheet in one component corresponds to a region of
macroscopic density in the other component. These features can be
seen in the density plots of Fig. \ref{03neg}(c).
%For these large values of rotation, many vortices are produced and the vortices in each component congregate into rows.

\begin{figure}
\begin{center}
\includegraphics[scale=0.5]{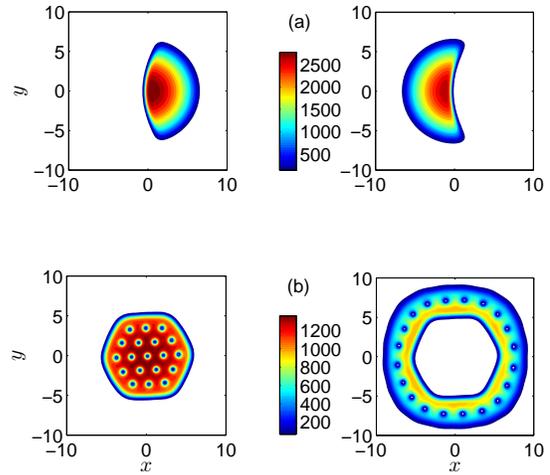}
\end{center}
% \caption{(Color online) A series of density plots for
% component-1 (left column) and component-2
%  (right column). The parameters are $g_1=0.0078$, $g_2=0.0083$ and $N_1=N_2=10^5$ and with
%   $\eta=\xi=1$ (set ES1) and $\Gamma_{12}=-1.3$ (which gives $g_{12}=0.0122$). The angular velocity of rotation is $\Omega$ and it takes the values (a) $0$, (b) $0.1$, and (c) $0.9$. In (a) and (b) the components are rotating droplets (region 4 of Fig. \ref{sch_2}) while in (c) the components have spatially separated (but are not symmetry broken) and there are no isolated density peaks (region 3 of Fig. \ref{sch_2}). Distance is measured in units of $r_0$ and density in
% units of $r_0^{-2}$.}
\caption{(Color online) A series of density plots for
component-1 (left column) and component-2
 (right column). The parameters are $g_1=0.0078$, $g_2=0.0083$ and $N_1=N_2=10^5$ and with
  $\eta=\xi=1$ (set ES1) and $\Gamma_{12}=-1.3$ (which gives $g_{12}=0.0122$). The angular velocity of rotation is $\Omega$ and it takes the values (a) $0$ and (b) $0.9$. In (a) the components are rotating droplets (region 4 of Fig. \ref{sch_2}) while in (b) the components have spatially separated (but keep some symmetry) and there are no isolated density peaks (region 3 of Fig. \ref{sch_2}). Distance is measured in units of $r_0$ and density in
units of $r_0^{-2}$.}
\label{13neg}
\end{figure}

{\em Region 3}: Spatial separation preserving some symmetry. Here
one component is a disk while the other component is an annulus and the disk
 fits within the annulus with a
boundary layer region in which both components have microscopically
small density as illustrated in Figure \ref{03neg}(b) and \ref{13neg}(b). Under rotation, the topological defects can
either be coreless vortex lattices  in the disk and (but not always)
corresponding isolated density peaks in the annulus, and/or
 a giant skyrmion at the boundary interface between the two
components, as will be described later.

\begin{figure}
\begin{center}
\includegraphics[scale=0.5]{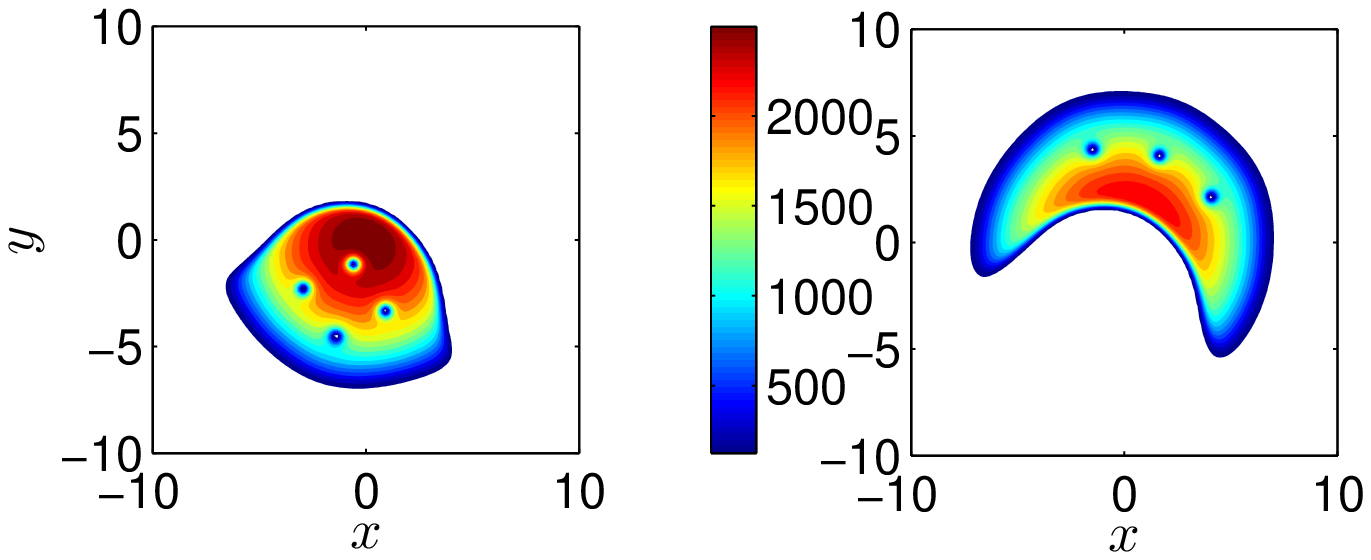}
\end{center}
\caption{(Color online) A density plot for component-1 (left column) and component-2 (right column) showing examples of vortex nucleation in rotating droplets. The parameters are $g_1=0.0078$, $g_2=0.0083$ and $N_1=N_2=10^5$ and with $\eta=\xi=1$ (set ES1) and $\Gamma_{12}=-9$ and $\Omega=0.5$. At these parameters the components are in region 4 of the phase diagram of Fig. \ref{sch_2}. Distance is measured in units of $r_0$ and density in
units of $r_0^{-2}$.}
\label{drop}
\end{figure}

{\it Region 4}: Rotating droplets. The components spatially separate
and have broken symmetry such that the centres of each condensate
are different and each component contains a single patch of density
 as illustrated in Figures  \ref{03neg}(a) and \ref{13neg}(a). In the rotating droplets, vortices can appear provided the
rotation is greater than some critical value. These features can be seen in the density plots
of Fig. \ref{drop}.

%\subsection*{D. Numerical $\Gamma_{12}-\Omega$ Phase Diagrams}

\begin{figure}
\begin{center}
\includegraphics[scale=0.4]{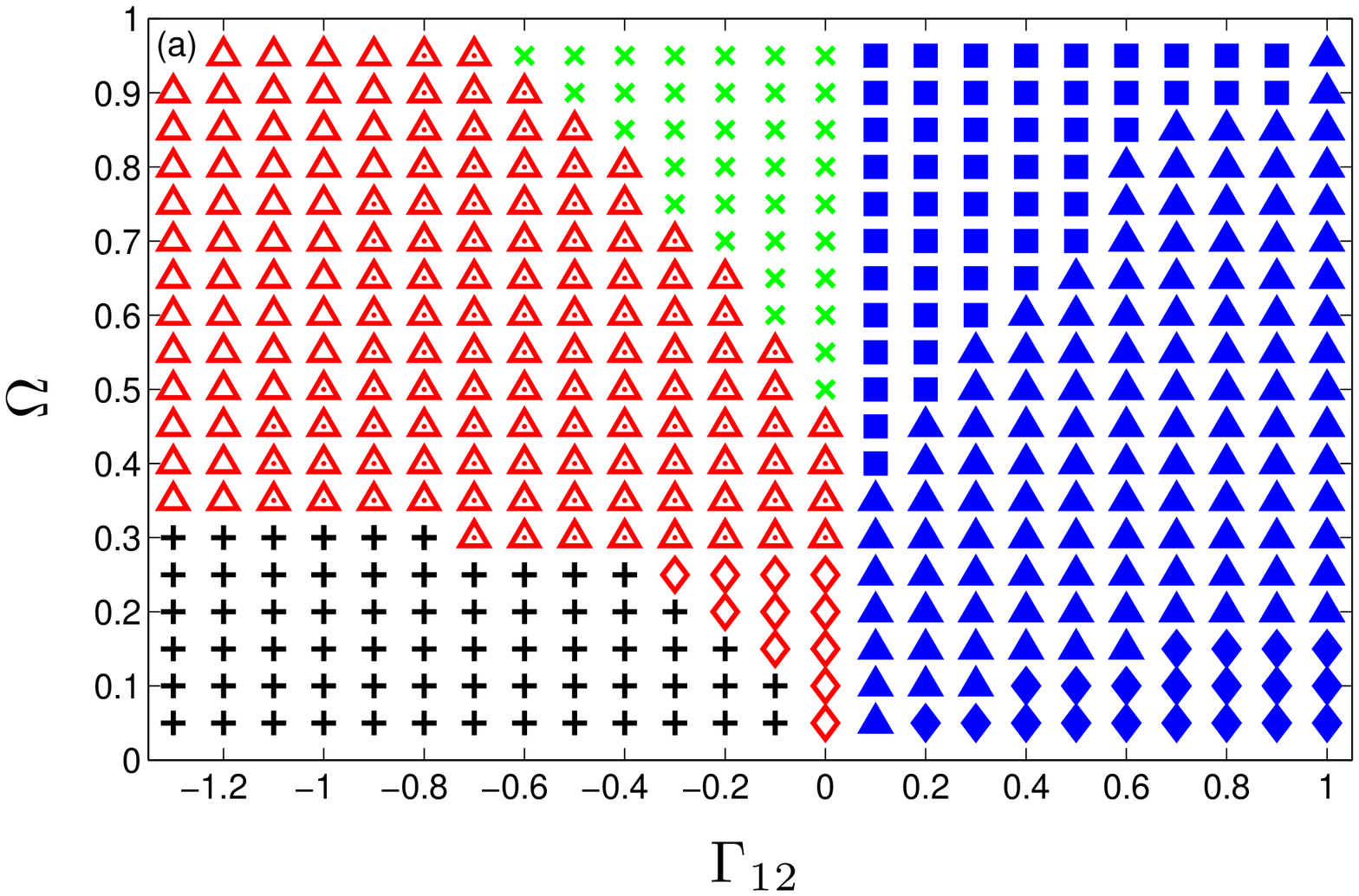}\\
\includegraphics[scale=0.4]{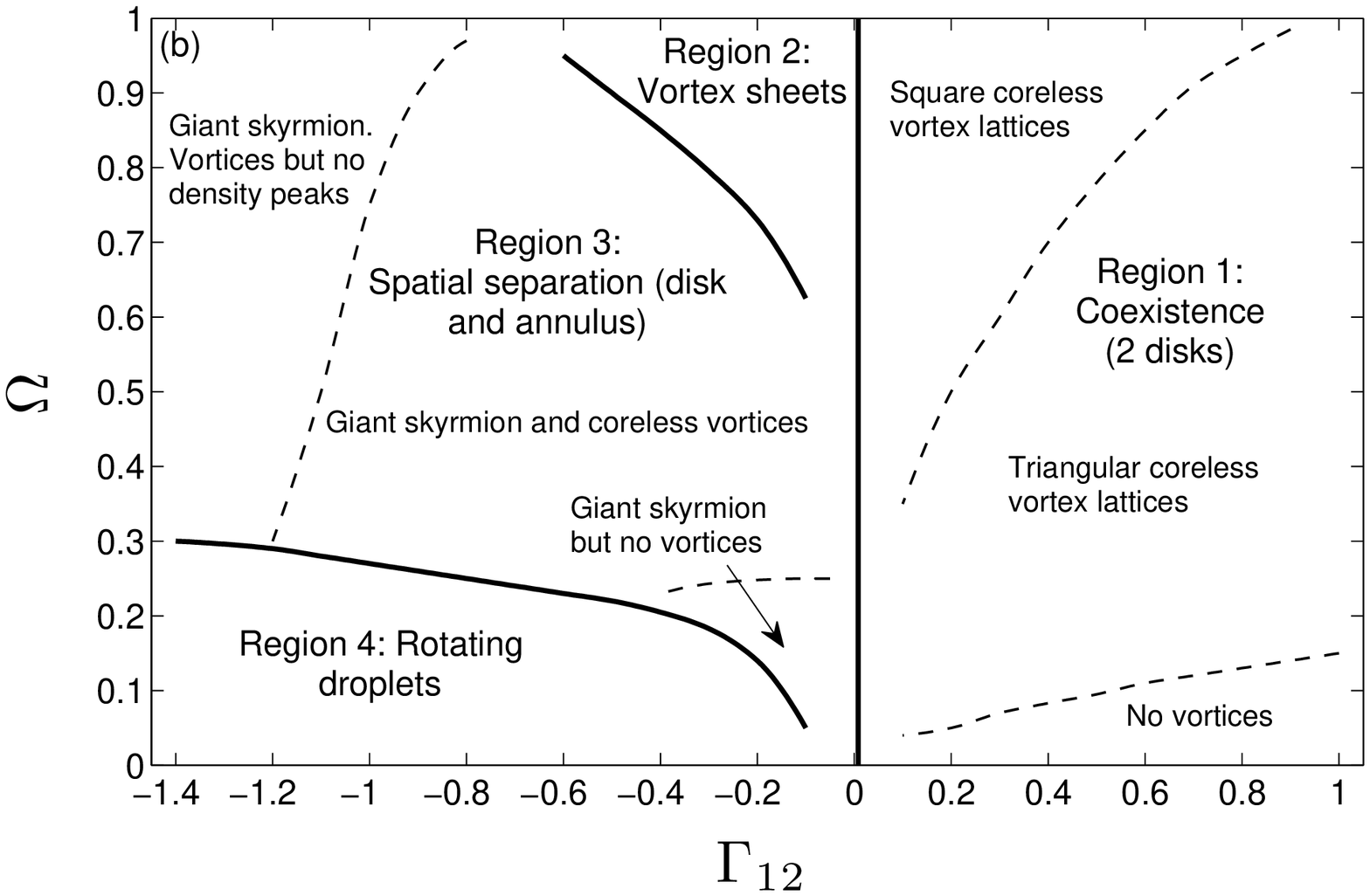}
\end{center}
\caption{(Color online) $\Omega-\Gamma_{12}$ phase diagrams for parameters
$g_1=0.0078$, $g_2=0.0083$, $N_1=N_2=10^5$ with $\eta=\xi=1$ (set ES1)
with normalisation taken over the individual components (Eq.
(\ref{norm})). (a) Numerical simulations where triangles (squares) indicate
that the vortex lattice in both components is triangular (square) and diamonds that no vortices
have been nucleated. Filled triangles,
squares and diamonds are where the two components are disk-shaped
and coexist. The empty triangles and empty diamonds
are where the two components have spatially separated: one component
is a disk and the other an annulus with a giant skyrmion at the
boundary layer; those triangles with a dot in the centre
represent the appearance of coreless vortices inside the annulus.
The crosses `x' are where the two components
have broken symmetry and are vortex sheets and the crosses `+' are rotating droplets.
%The vertical dotted lines represent the maximum value of $\Gamma_{12}=1$ ($g_{12}=0$), when $g_{12}=g_1$ ($\Gamma_{12}=0.13$), when $g_{12}^2=g_1g_2$ ($\Gamma_{12}=0$) and when $g_{12}=g_2$ ($\Gamma_{12}=-0.15$).
(b) A schematic representation of the numerical simulations. The
solid lines indicate the boundary between different identified
regions (determined by the geometry of the ground state)
and the dashed lines the boundary between triangular and
square lattices in region 1 and the boundary between peaks and no
peaks in region 3. The boundary between region 1 and the others can
be calculated analytically by Eq. (\ref{spatsep}) to give
$\Gamma_{12}=0.008$. The unit of rotation is $\tilde{\omega}$.}\label{sch_2}
\end{figure}

\begin{figure}
\begin{center}
\includegraphics[scale=0.4]{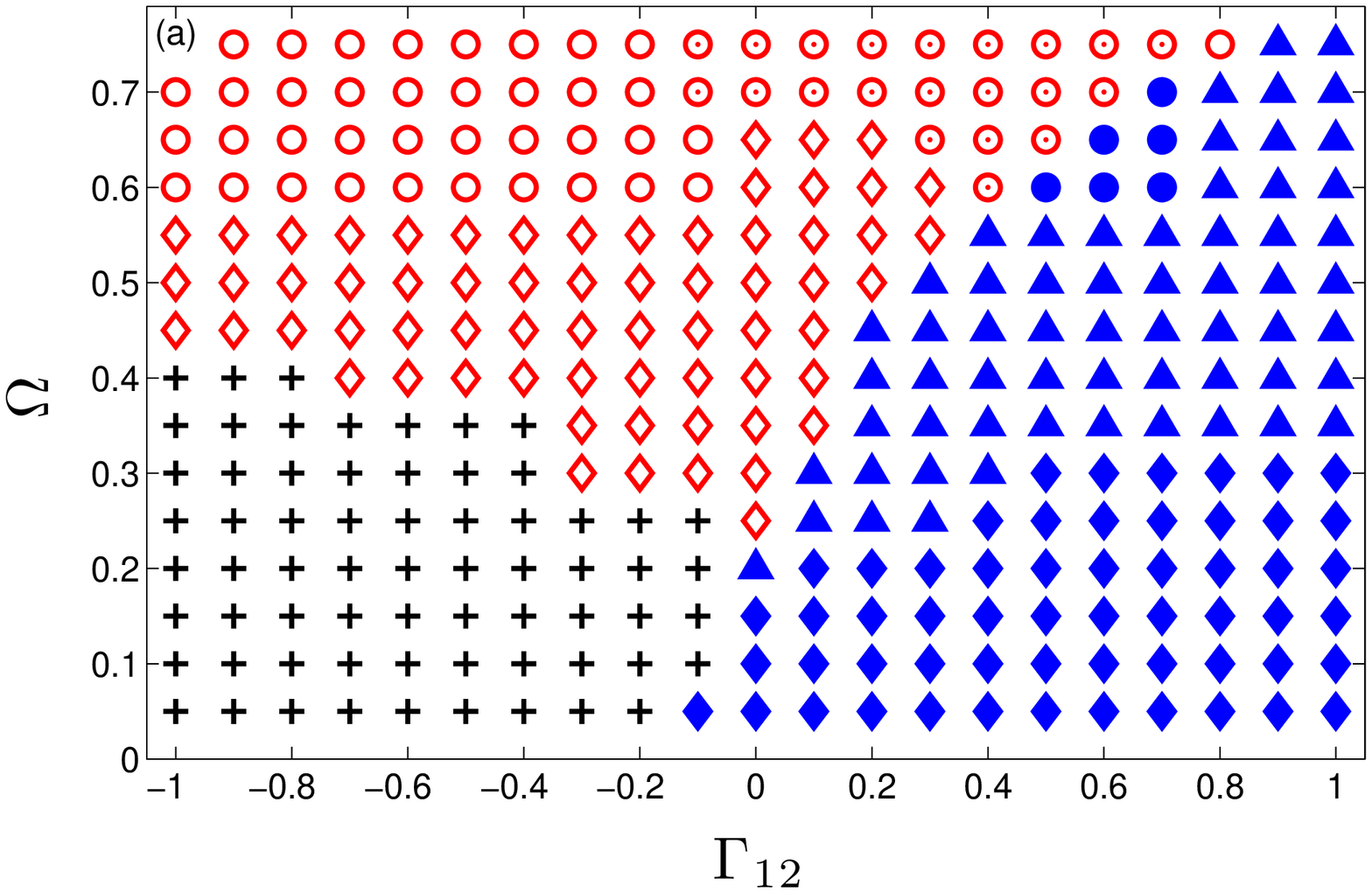}\\
\includegraphics[scale=0.4]{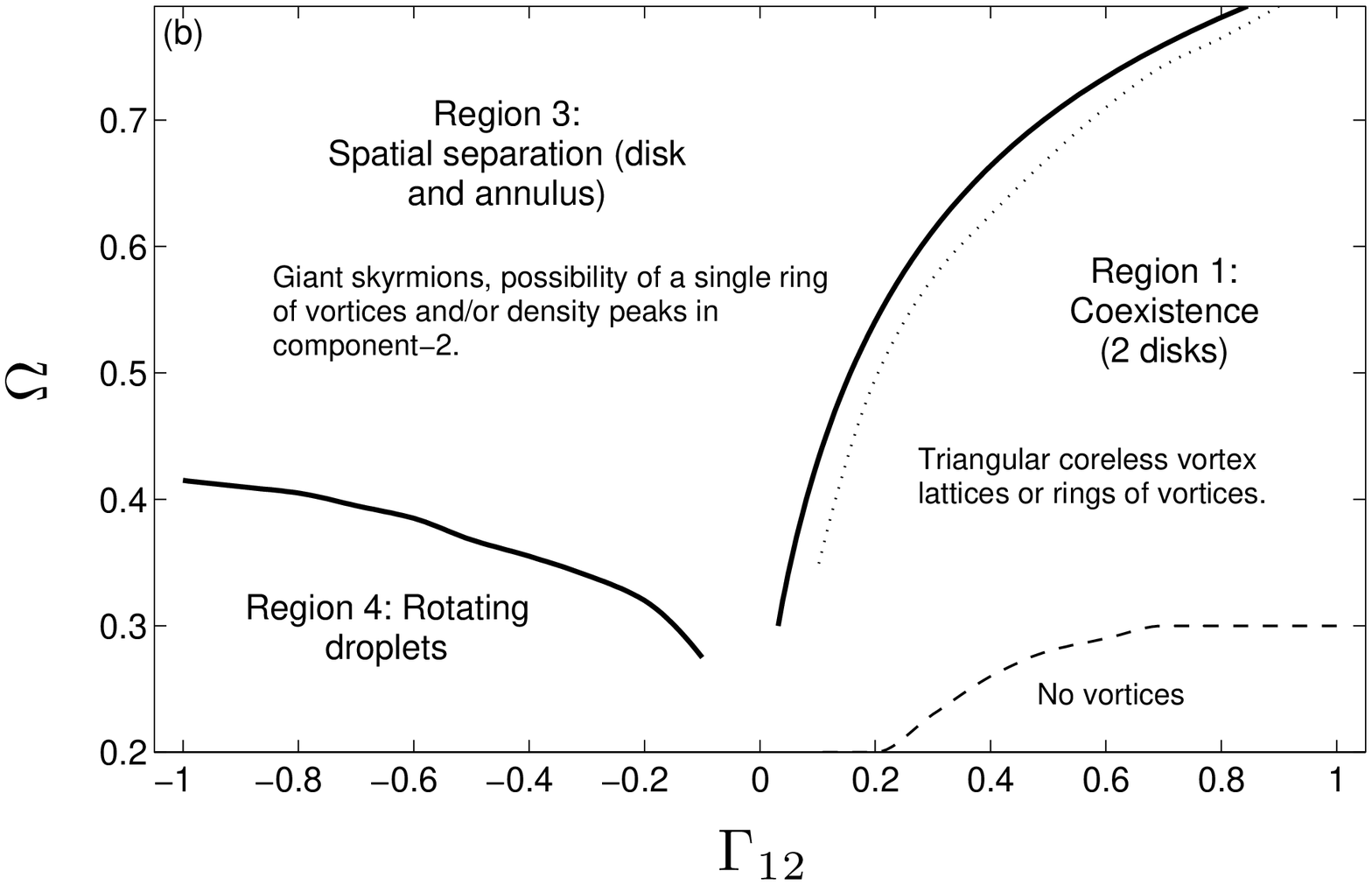}
\end{center}
\caption{(Color online) $\Omega-\Gamma_{12}$ phase diagrams for parameters $g_1=0.0067$, $g_2=0.0063$, $N_1=N_2=10^4$ with $\eta=0.48$ and $\eta\xi^2=1$ (set ES2) with normalisation taken over the individual components (Eq. (\ref{norm})). (a) Numerical simulations where triangles indicate that the vortex lattice in component-2 is triangular, circles that component-2 contains rings of vortices and diamonds where no vortices have been nucleated. Filled triangles, circles and diamonds are where the two components are disk-shaped and coexist. The empty circles and empty diamonds are where the two components have spatially separated: one component is a disk and the other an annulus with a giant skyrmion at the boundary layer; those circles with a dot in the centre represent the appearance of coreless vortices inside the annulus. The crosses `+' are where the two components have broken symmetry and are rotating droplets.
%The vertical dotted lines represent the maximum value of $\Gamma_{12}=1$ ($g_{12}=0$), when $g_{12}=g_1$ ($\Gamma_{12}=0.27$), when $g_{12}^2=g_1g_2$ ($\Gamma_{12}=0$) and when $g_{12}=g_2$ ($\Gamma_{12}=-0.36$).
(b) A schematic representation of the numerical simulations.
The solid lines indicate the boundary between different identified regions (determined by the geometry of the ground state).
The boundary between region 1 and region 3 is also calculated analytically by Eq. (\ref{spatsep}) (dotted line).
For these parameters there is no region 2. The unit of rotation is $\tilde{\omega}$.}\label{sch_3}
\end{figure}

\begin{figure}
\begin{center}
\includegraphics[scale=0.4]{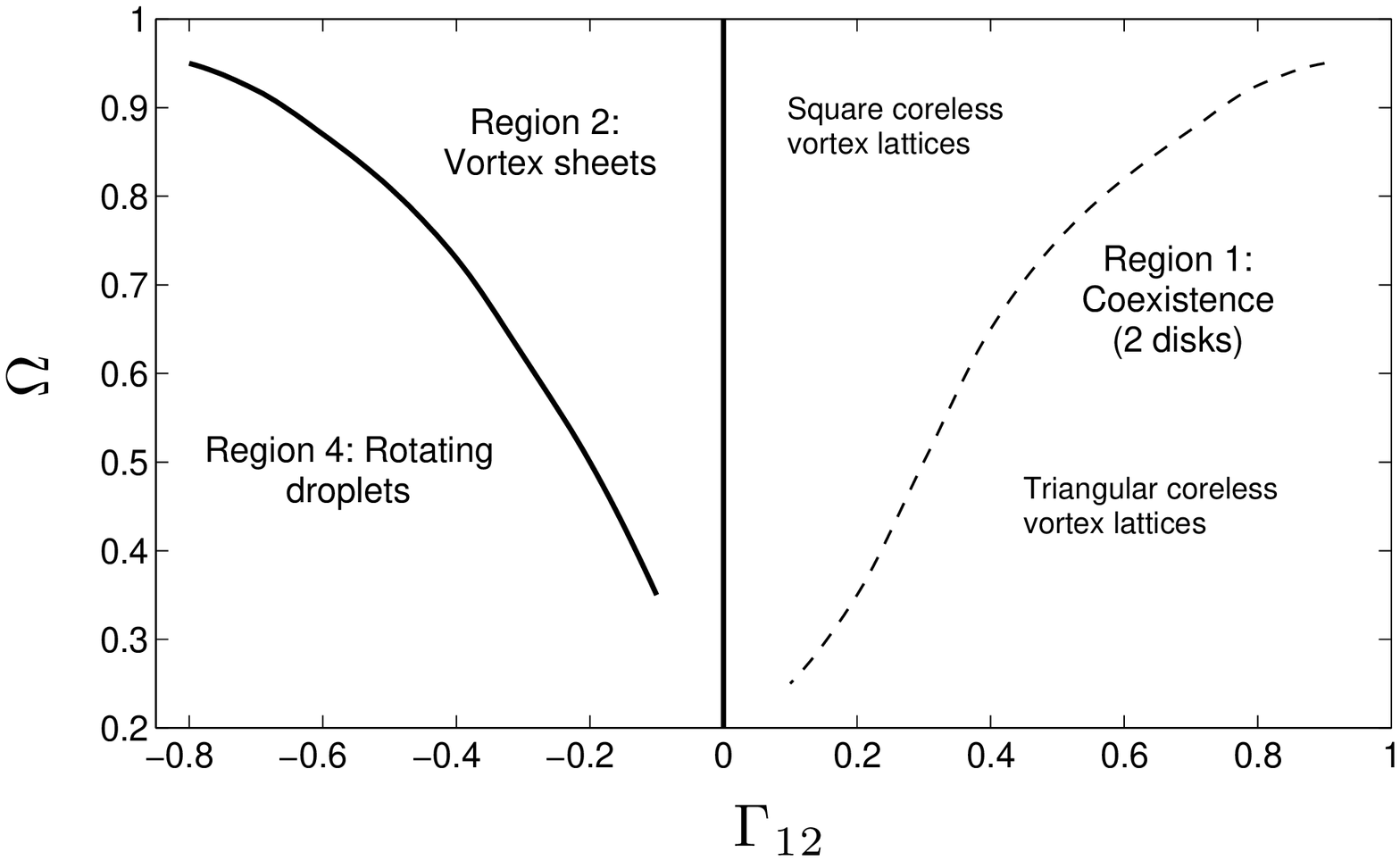}
\end{center}
\caption{A schematic $\Omega-\Gamma_{12}$ phase diagram for $g_1=g_2\equiv g=0.0078$, $N_1=N_2=10^5$ where $\eta=\xi=1$ (set ES3) and with the normalisation taken over the individual components (Eq. (\ref{norm})). The solid lines indicate the boundary between different identified regions (determined by the geometry of the ground state) and the dashed lines the boundary between triangular and square lattices. For these parameters there is no region 3. The unit of rotation is $\tilde{\omega}$.}\label{sch_1}
\end{figure}

The $\Omega-\Gamma_{12}$ phase diagrams corresponding to the
experimental parameter sets introduced above (sets ES1, ES2 and ES3)
are shown in Fig.'s \ref{sch_2}, \ref{sch_3}  and \ref{sch_1}.  The
last one is of similar type as the one reported by Kasamatsu et al \cite{ktu1}
(there $g_1=g_2=4000$ and $N_1=N_2=1/2$).

New features can be observed in Fig \ref{sch_2} and \ref{sch_3},
such as isolated density peaks that eventually vanish as
$\Gamma_{12}$ is made more negative and the multiply quantised
skyrmions at the interface between the disk component and the
annular component in region 3. When $\{\eta,\xi\}\neq1$ (Fig. \ref{sch_3}), no region 2
is found to exist.  This absence is easily explained by two factors:
the onset of region 3 for (large) positive values of $\Gamma_{12}$
and the lack of vortices nucleated in component-1 (and to some
extent in component-2).

We will now analyze in more detail some of
the features of regions 1, 3 and 4 of the phase diagrams.  Vortex
sheets (region 2) have been analysed (mainly in the case of equal $g_k$, $m_k$
and $\omega_k$) in \cite{kt}.

\subsection*{C. Analysis of the Features of the Phase Diagrams}

\subsubsection*{1. Square lattices}

It is a specific property of two components to stabilize square
lattices. A requirement for the existence of square lattices is the
nucleation of many vortices in both components, something not
permitted in ES2 (Fig. \ref{sch_3}), where only a small number of vortices
are ever nucleated in component 1.

Square lattices generally occur at high rotational velocities and
examples have been observed experimentally \cite{sch} and
numerically \cite{wcbb,ktu,ktu1}. Mueller and Ho \cite{mh} and later
Ke{\c c}eli and \"Oktel \cite{ko} have analysed the transition from
triangular to square lattices as $\Gamma_{12}$ is varied in
two-component condensates when $g_1\sim g_2$ and $\Omega$ is such that the condensate is in the lowest Landau
level (LLL). Providing $\Omega$ is large, according to \cite{ko,mh}
the value of $\alpha=\sqrt{1-\Gamma_{12}}$ determines whether the
vortex lattices are triangular ($0<\alpha<0.172$), in transition to
becoming square ($0.172<\alpha<0.373$) or square ($\alpha>0.373$).
Therefore, assuming that, in the example parameters of $g_1=0.0078$,
$g_2=0.0083$ and $\eta=\xi=1$ (set ES1), the $g_k$ are sufficiently close
enough, the square lattice should first appear at $\alpha=0.37$, or
equivalently $\Gamma_{12}=0.86$. Comparing this to the numerically
obtained values, where the square lattice first appears for
$0.8<\Gamma_{12}<0.9$ in Fig. \ref{sch_2}, the agreement seems good.
 It will be interesting to see how the
analysis of \cite{ko,mh} has to be modified when $g_1$ becomes distant from $g_2$.

Nevertheless, the square lattices are present for lower rotational
velocities at which the LLL is not necessarily valid. In a
forthcoming paper, using the nonlinear sigma model
presented below, we hope to derive a vortex energy
 in terms
of the positions of vortices. The ground state of this energy indeed
leads to square lattices in some range of parameters (see also Section III.E). It turns out
that this vortex energy is similar to that of Barnett et al
\cite{barnett,barnett2}.

\subsubsection*{2. Rotating droplets}

In the parameter range of region 4, symmetry breaking  with spatial
separation occurs. When the condensate is not under rotation, the
ground states of the two components will be `half ball'-like
structures, as can be seen in Fig. \ref{13neg}(a) (if $g_1=g_2$ and
$\eta=\xi=1$, then the two components are exactly half-balls). The
difference between the intercomponent strengths introduces a
curvature to the inner boundary of each component with the component
that corresponds to larger $g_k$ having positive curvature and the
other component having negative curvature. This curvature depends on
both $\Gamma_{12}$ and $\Omega$. If $\Gamma_{12}$ is held constant,
then the curvature increases as $\Omega$ increases. When
$\Gamma_{12}$ goes to $-\infty$, the droplets approach half-balls.

Vortex nucleation is also seen in region 4; see Fig. \ref{drop}. In this figure there are four vortices in component-1 and three vortices in component-2. The number of vortices in each component will increase as $\Omega$ is increased but this increase will also increase the curvature of the inner boundaries of the components, thus preventing the vortices aligning themselves into vortex sheets.
%Note that the vortices that are created in these rotating droplets do not have corresponding isolated density peaks.
Examples of the rotating droplet ground states can be seen in Fig. \ref{13neg}(a), and further examples can be found in  \cite{kyt,ohberg,ohberg1,navarro,chris}.

\subsubsection*{3. Spatial separation preserving some symmetry: disk plus annulus}

%So far the features of the phase diagram for $g_1\neq g_2$ have direct parallels with the phase diagram for $g_1=g_2$; in region 1 the condensates are both disks and the coreless vortices are arranged onto triangular or square lattices depending on the values of $\Gamma_{12}$ and $\Omega$ and in region 2 vortex sheets are present. However, as $\Gamma_{12}$ is decreased, the effects caused by the difference between $g_1$ and $g_2$ become more prominent. At $\Gamma_{12}\sim\epsilon$ with $\epsilon>0$, the two components begin to spatially separate.
 Region 3 is defined by the $\Gamma_{12}$ and $\Omega$ in which the
condensates have spatially separated components but still possess some symmetry about the origin: the disk component-1 sits securely within the annular
component-2 (see Fig.'s \ref{03neg}(b) and \ref{13neg}(b)). %It exists when $\Lambda g_1\neq g_2$.
There is a boundary layer evidenced where the outer edge
of the disk overlaps the inner edge of the annulus.

 Let us describe the onset of region 3 from region 4, captured in Fig. \ref{03neg} (sub-image (a) to (b) or (c) to (b)). For
 a particular $\Gamma_{12}$,  as $\Omega$ is increased, then the curvature increases  to such an extent  that the components develop constant non-zero curvature, and are identified as a disk and an annulus.
% and in Fig. \ref{13neg} (sub-image (b) to (c)).
Conversely, if $\Omega$ is held constant and $\Gamma_{12}$ is pushed to more negative values, then the curvature decreases.
%, as can be viewed by comparing Fig.'s \ref{03neg}(a) and \ref{13neg}(b).

Under rotation, defects can be observed: coreless vortices and giant
skyrmions. Coreless vortices sit inside the disk while giant
 skyrmions are observed in the boundary layer.

 \subsubsection*{4. Coreless Vortices in the disk plus annulus case}

 In the geometry of disk plus annulus, the vortices in the inside disk have corresponding isolated
density peaks  inside the annulus, hence in the microscopic density
regions, as illustrated in Fig. \ref{03neg}(b). Thus any vortex
lattice structure produced in component-1 is replicated by a peak
lattice structure in component-2, i.e. there is the continuing
presence of coreless vortex lattices in the spatially separated
condensates \cite{note2}.

 As the number of
vortices in the disk  increases with increasing angular
velocity, the number of density peaks inside the central hole of the
annulus likewise increases. This has the effect that the central hole
of the annulus can be masked at high angular velocities by the
multiple occurrence of the density peaks.  A recent analytical
understanding of the interaction between vortices and peaks
 has been provided by \cite{kasa}. It may be  extended to the case of the disk and annulus, for which the average density in the central hole is very small for one component.
  
% \begin{figure}
% \begin{center}
% \includegraphics[scale=0.5]{peaks_3d.eps}
% \end{center}
% \caption{(Color online) 3D density plots for component-1 (left column) and component-2 (right column) displaying the isolated density plots in component-2. The parameters are $g_1=0.0078$, $g_2=0.0083$ and $N_1=N_2=10^5$ and with $\eta=\xi=1$ (set ES1) and $\Gamma_{12}=-0.3$ (which gives $g_{12}=0.0092$) and $\Omega=0.5$. At these parameters the components are in region 3 of the phase diagram of Fig. \ref{sch_2}. Distance is measured in units of $r_0$ and density in
% units of $r_0^{-2}$.}\label{peaks_3d}
% \end{figure}
Pushing $\Gamma_{12}$ to lower values has the effect of reducing the
size of the boundary layer between the disk and annular components,
but also the isolated density peaks disappear; see Fig.
\ref{13neg}(b). A recent work has analysed, from an energy perspective, the preference for the ground state to contain or not contain density peaks \cite{cate}.

\begin{figure}\begin{center}
\includegraphics[scale=0.5]{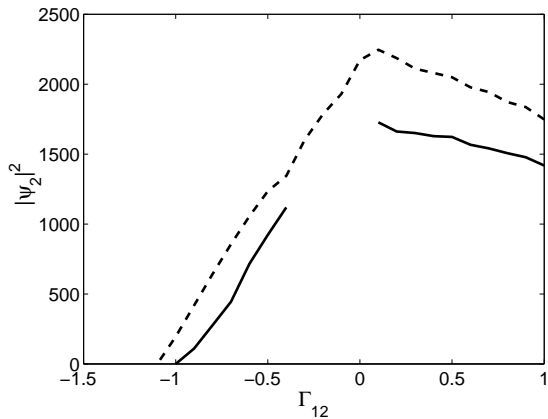}\end{center}
\caption{Plots of maximum peak density in component-2 against
$\Gamma_{12}$ for $\Omega=0.5$ (dashed line) and $\Omega=0.75$
(solid line) when $g_1=0.0078$, $g_2=0.0083$, $N_1=N_2=10^5$ and
with $\eta=\xi=1$ (set ES1) . The disjoint region when $\Omega=0.75$
for $-0.4<\Gamma_{12}<0.1$ is due to the absence of isolated density
peaks in component-2 as a result of the appearance of vortex sheets.
Density is measured in units of
$r_0^{-2}$.}\label{peaks}\end{figure}

 Figure \ref{peaks} plots the maximum peak density of
the density peaks in component-2 as a function of $\Gamma_{12}$ for
the parameters of Fig. \ref{sch_2} and with $\Omega=0.5$ and
$\Omega=0.75$.  The disjoint region when $\Omega=0.75$ for
$-0.4<\Gamma_{12}<0.1$ on Fig. \ref{peaks} is due to the absence of
density peaks in component-2 as a result of the appearance of vortex
sheets (region 4). We see from Fig. \ref{peaks},
 that the maximum peak density  occurs when
$\Gamma_{12}\sim0.01$ which is the value at which the components
begin to spatially separate (the transition between region 1 and
region 3) in set ES1.  For $\Gamma_{12}$ and $\Omega$ that take
values outside of region 1, the maximum density of the peaks
decreases linearly as $\Gamma_{12}$ increases. The maximum density approaches microscopically small values for
$\Gamma_{12}\sim-1.1$ when $\Omega=0.5$ and for $\Gamma_{12}\sim-1$
when $\Omega=0.75$. An example of the transition can be seen in Fig.'s
\ref{03neg}(b) and \ref{13neg}(b).
%While the density peaks have
%disappeared from the density profiles, the signatures of the
%coreless vortices are still present in a phase profile plot.

\subsubsection*{5. Giant skyrmion}

A boundary layer between the overlap of component-1 and component-2
 is present for all values of $\Gamma_{12}$ in region 3 but reduces
in width as $\Gamma_{12}$ becomes more negative (indeed the boundary
layer disappears only as $\Gamma_{12}\rightarrow-\infty$). There are
additional topological defects that occur in the boundary layer that
are not discernible on a traditional density plot. These
topological defects' presence can be observed in a phase profile,
however a better visualisation is to use the pseudospin
representation (\ref{rhot})-(\ref{seqs}).
%(\ref{rhot})-(\ref{sx})-(\ref{sy})-(\ref{sz}).
    One can plot the functions $S_x$, $S_y$ and/or $S_z$
    which reveal the presence of all the topological defects -
     the coreless vortices (singly quantised skyrmions) that have already been visualised on the plots,
      and the new defect, a multiply quantised skyrmion.

\begin{figure}
\begin{center}
\includegraphics[scale=0.5]{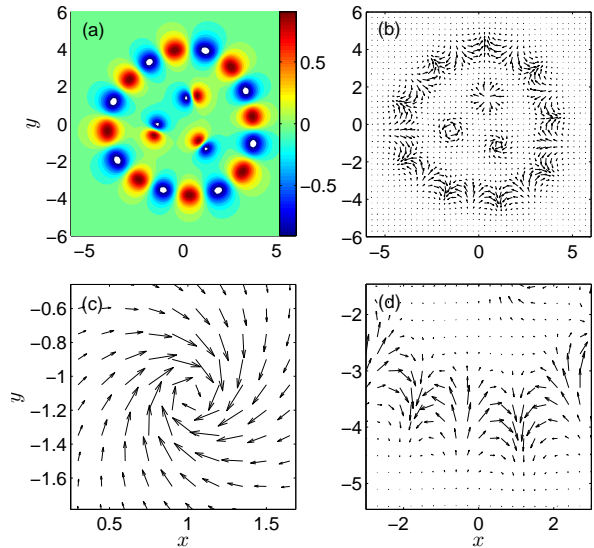}
\end{center}
\caption{(Color online) Plots of (a) $S_x$ and (b) a vectorial plot of $(S_x,S_y)$ for $\Gamma_{12}=-0.3$ and $\Omega=0.5$ when $g_1=0.0078$, $g_2=0.0083$, $N_1=N_2=10^5$ and with $\eta=\xi=1$ (set ES1). The ring of skyrmions, at $r=r_s$, traces the boundary layer and the coreless vortices exist for $r<r_s$. Blow-ups (c) and (d) of (b) highlight the nature of the two types of topological defect; (c) a coreless vortex at $(1,-1)$ and (d) a section of the multiply quantised skyrmion. The respective density plots for this case are shown in Fig. \ref{03neg}(b). Distance is measured in units of $r_0$.}
\label{topdef}
\end{figure}

The distinct nature of the two types of topological defect can be
clearly seen in Fig. \ref{topdef}(a,b) where an $S_x$ plot and an
$(S_x,S_y)$ vectorial plot are shown  respectively. A density plot of each component for these same parameters is seen in
Fig. \ref{03neg}(b). The coreless vortices evident in Fig.
\ref{03neg}(b) are again clearly evident in Fig. \ref{topdef}(a) and
(b). A blow-up of the region close to one of (the three) coreless
vortices in the $(S_x,S_y)$ vectorial plot, centred at $(1,-1)$ and
exhibiting circular disgyration is shown in Fig. \ref{topdef}(c).
The texture of $\bm{S}$ can also exhibit cross- and radial-
disgyration \cite{ohmi}. Conversely, the multiply quantised
skyrmion, not present in the density plots of Fig.
\ref{03neg}(b), can clearly be seen in both Fig. \ref{topdef}(a) and
(b). The multiply quantised skyrmion forms a ring along the boundary
layer. A blow-up around $(0,-3.5)$ for the $(S_x,S_y)$ vectorial
plot again shows the multiply quantised nature of this defect. A
multiply quantised skyrmion was evidenced by \cite{ywzf} who termed
it a giant skyrmion, a terminology retained in this paper.
 Increasing the rotation results in an increase in both the number of
coreless vortices and in the multiplicity of the giant skyrmion.
Ref \cite{ywzf} gave a relationship, $m=n+1$, between the
multiplicity of the giant skyrmion $n$ and the total circulation $m$
in the central hole of component-2. However the numerical
simulations here suggest that a more appropriate correlation is
provided by $m=n+l$ where $l$ is the number of isolated coreless
vortices. As an example, consider Fig. \ref{topdef} where
$\Omega=0.5$ and $n=8$, $l=3$ and $m=11$, satisfying $m=n+l$. It
should be noted that the relationship $m=n+l$ still holds as the
isolated coreless density peaks disappear from the density
profile of component-2; this is because the phase imprints due to the vortices in component-1,
which
constitute $l$, are still present.

\section*{III. A Non-Linear $\sigma$-Model}
\label{tfprof}

The ground states of the rotating two-component
condensate can be recovered from a `non-linear $\sigma$-model'
formulation of the energy functional in terms of the total density
$\rho_T$ and spin vector $\bf S$ (Eq.'s
(\ref{rhot})-(\ref{seqs})). We will analyze in detail the possible regimes in this section and show
how a Thomas Fermi approximation can hold in the case $\Gamma_{12}>0$
 or can be generalized and provide relevant information in the case $\Gamma_{12}<0$.

\subsection*{A. Energy Functional Representation}

We write the energy functional of the two wave functions  (Eq.
(\ref{enfn}), non-dimensionalised) as
$E[\psi_1,\psi_2]=E_{\text{KE}}+E_{\text{PE}}+E_{\text{I}}$ where
\begin{subequations}
    \label{energyps}
    \begin{align}
   \begin{split} E_{\text{KE}}&=\frac{1}{\eta+1}\int\left|\left(\nabla-\frac{i}{2}(\eta+1)\bm{\Omega}\times\bm{r}\right)\psi_1\right|^2\\
    &\quad+\eta\left|\left(\nabla-\frac{i}{2\eta}(\eta+1)\bm{\Omega}\times\bm{r}\right)\psi_2\right|^2\quad d^2r,
\end{split}\\
    E_{\text{PE}}&=\int2j_1r^2|\psi_1|^2+2j_2r^2|\psi_2|^2\quad d^2r,\\
    E_{\text{I}}&=\int\frac{1}{2}g_1|\psi_1|^4+\frac{1}{2}g_2|\psi_2|^4+g_{12}|\psi_1|^2|\psi_2|^2\quad d^2r.
\end{align}
\end{subequations}
with
\begin{subequations}
\begin{align}
    j_1=&\frac{1}{2}\frac{(1+\eta)\xi^2}{(1+\xi)^2}-\frac{1}{8}(1+\eta)\Omega^2,\\                  j_2=&\frac{1}{2}\frac{(1+\eta)}{\eta(1+\xi)^2}-\frac{1}{8\eta}(1+\eta)\Omega^2.
\end{align}
\end{subequations}
It is assumed that the $g_k$, $m_k$, $\omega_k$ and $N_k$ are distinct
%with $g_2\ge \Lambda g_1$,
so that a weighted total density can be defined as (\ref{rhot}) with
 $ \psi_1=\sqrt{\rho_T}\chi_1$ and $
\psi_2=\sqrt{\rho_T/\eta}\chi_2$. The spin density vector $\bm{S}$,
which satisfies $|\bm{S}|^2=1$ everywhere, has components given by
Eq.'s (\ref{seqs}). We then have
\begin{equation}
    |\psi_1|^2=\frac{1}{2}\rho_T(1+S_z),\quad|\psi_2|^2=\frac{1}{2\eta}\rho_T(1-S_z).
\end{equation}
We introduce the phases $\theta_1$ and $\theta_2$ defined by $\chi_1=|\chi_1|\exp(i\theta_1)$ and $\chi_2=|\chi_2|\exp(i\theta_2)$.
 The energy functional is
expressed in terms of 4 variables: the total density $\rho_T$, the component $S_z$
 and the angles $\theta_1$, $\theta_2$. We see that $E_{\text{KE}}= E_{\rho_T}+E_{\text{Sz}}+E_{\theta_1,\theta_2}$ where
\begin{subequations}
    \begin{align}
        E_{\rho_T}=&\int\frac{1}{(\eta+1)}(\nabla\sqrt{\rho_T})^2\quad d^2r,\\
        E_{\text{Sz}}=&\int \frac{1}{4}\frac{\rho_T}{(\eta+1)}\frac{(\nabla{{S_z}})^2}{(1-S_z^2)}\quad d^2r,\\
        E_{\theta_1,\theta_2}=&\int \frac{1}{2}\frac{\rho_T}{(1+\eta)}\Bigg[(1+S_z)\left(\nabla\theta_1-\frac{1}{2}(1+\eta)\bm{\Omega}\times\bm{r}\right)^2\nonumber\\
        &\quad+(1-S_z)\left(\nabla\theta_2-\frac{1}{2\eta}(1+\eta)\bm{\Omega}\times\bm{r}\right)^2\Bigg]\quad d^2r.
    \end{align}
\end{subequations}
%Rewriting $E_{\theta_1,\theta_2}$ using
%\begin{subequations}
%\begin{align}
%    \nabla\theta_1=&\frac{1}{2}\left(\nabla\Theta+\bm{R}/(1-S_z^2)\right),\\
%\nabla\theta_2=&\frac{1}{2}\left(\nabla\Theta-\bm{R}/(1-S_z^2)\right),
%\end{align}
%\end{subequations}
%where $\bm{R}=S_y\nabla S_x-S_x\nabla S_y$ and $\Theta=\theta_1+\theta_2$, gives (relabelling $E_{\theta_1,\theta_2}$ by $E_{\Theta}$)
%\begin{equation}
%    \begin{split}
%&E_{\Theta}=\int\frac{\rho_T}{2(1+\eta)}\times\\
% &\Bigg[(1+S_z)\left(\frac{1}{2}\nabla\Theta+\frac{\bm{R}}{2(1-S_z^2)}-\frac{1}{2}(1+\eta)\bm{\Omega}\times\bm{r}\right)^2\nonumber\\        &+(1-S_z)\left(\frac{1}{2}\nabla\Theta-\frac{\bm{R}}{2(1-S_z^2)}-\frac{1}{2\eta}(1+\eta)\bm{\Omega}\times\bm{r}\right)^2\Bigg]\quad d^2r.
%&\Bigg[(1+S_z)\left(v_{\text{eff(1)}}-\frac{1}{2}(1+\eta)\bm{\Omega}\times\bm{r}\right)^2\\
%&\quad+(1-S_z)\left(v_{\text{eff(2)}}-\frac{1}{2\eta}(1+\eta)\bm{\Omega}\times\bm{r}\right)^2\Bigg]\quad d^2r.
%\end{split}
%\end{equation}
%Note that under this terminology, $\nabla\theta_1-\nabla\theta_2=\bm{R}/(1-S_z^2)$. This expression
%yields two different effective velocities for the two different components.%: $v_{\text{eff(1)}}=$ and $v_{\text{eff(2)}}=$.
The other terms of the energy straightforwardly become
\begin{subequations}
    \begin{eqnarray} E_{\text{PE}}=&&\int\left[(j_1+j_2/\eta)+(j_1-j_2/\eta)S_z\right]r^2\rho_T\quad d^2r,\\     E_{\text{I}}=&&\int\frac{\rho_T^2}{2}\left(\bar{c}_0+\bar{c}_1S_z+\bar{c}_2S_z^2\right)\quad d^2r,
    \end{eqnarray}
\end{subequations}
with
\begin{subequations}
\begin{align}
\label{c00}
\bar{c}_0=&\frac{1}{4\eta^2}(\eta^2g_1+g_2+2\eta g_{12}),\\
    \bar{c}_1=&\frac{1}{2\eta^2}(\eta^2g_1-g_2),\\
\label{c22}
    \bar{c}_2=&\frac{1}{4\eta^2}(\eta^2g_1+g_2-2\eta g_{12}).
\end{align}
\end{subequations}

Thus the complete energy is
\begin{equation}
    \label{sig}
\begin{split}
E&=\int \frac{1}{(\eta+1)}(\nabla\sqrt{\rho_T})^2+\frac{\rho_T}{4(\eta+1)}\frac{(\nabla{{S_z}})^2}{(1-S_z^2)}\\
&+\frac{\rho_T}{2(1+\eta)}\times\\
%\Bigg[(1+S_z)\left(v_{\text{eff(1)}}-\frac{1}{2}(1+\eta)\bm{\Omega}\times\bm{r}\right)^2\\
%&\qquad\quad+(1-S_z)\left(v_{\text{eff(2)}}-\frac{1}{2\eta}(1+\eta)\bm{\Omega}\times\bm{r}\right)^2\Bigg]\\
&\Bigg[(1+S_z)\left(\nabla\theta_1-\frac{1}{2}(1+\eta)\bm{\Omega}\times\bm{r}\right)^2\\
&+(1-S_z)\left(\nabla\theta_2-\frac{1}{2\eta}(1+\eta)\bm{\Omega}\times\bm{r}\right)^2\Bigg]\\
&+\left[(j_1+j_2/\eta)+(j_1-j_2/\eta)S_z\right]r^2\rho_T\\
&+\frac{\rho_T^2}{2}\left(\bar{c}_0+\bar{c}_1S_z+\bar{c}_2S_z^2\right)\quad
d^2r.
\end{split}
\end{equation}
Energy (\ref{sig}) is subject to the constraints (\ref{norm}) that can be rewritten
as \begin{subequations}
    \label{nn2}
\begin{align}
    \label{nn2a}
    \int\rho_T(1-S_z)/2\quad d^2r=&N_2\eta,\\
    \int\rho_T(1+S_z)/2\quad d^2r=&N_1.
    \label{nn2b}
\end{align}
\end{subequations}or equivalently
\begin{subequations}
    \label{nn1}
\begin{align}
    \label{nn1a}
    \int\rho_T\quad d^2r=&N_1+N_2\eta,\\
    \int\rho_TS_z\quad d^2r=&N_1-N_2\eta,
    \label{nn1b}
\end{align}
\end{subequations}
The minimization of the energy under the constraints (\ref{nn1})
yields two coupled equations with two Lagrange multipliers, $\mu$ and $\lambda$. We will
write them
 when the phase is constant (the gradient of the phases
 $\theta_1$ and $\theta_2$ can be ignored) and the effect of rotation is contained in an effective centrifugal dilation of the trapping potential included in $j_1,j_2$:
\begin{equation}
    \begin{split}
    \label{muu1}
    \mu+\lambda S_z=&-\frac 1{(1+\eta)}\frac{\Delta \sqrt{\rho_T}}{\sqrt{\rho_T}}
    +\frac 1{4(1+\eta)}\frac{(\nabla S_z)^2}{(1-S_z^2)}\\&\quad+\left[(j_1+j_2/\eta)+(j_1-j_2/\eta)S_z\right]r^2\\
&\qquad+\rho_T(\bar{c}_0+\bar{c}_1S_z+\bar{c}_2S_z^2),
\end{split}
\end{equation}
and
%The Euler-Lagrange equation satisfying $S_z$ reduces to
\begin{equation}
    \begin{split}
    \label{szeq1}
    \lambda=&-\frac 1{4(1+\eta)}\frac{\Delta S_z+\nabla \rho_T \cdot\nabla S_z}{(1-S_z^2)}+\frac 1{2(1+\eta)}\frac{S_z(\nabla S_z)^2}{(1-S_z^2)}\\
&\quad +(j_1-j_2/\eta)r^2+\left({\bar{c}_1}+2\bar{c}_2S_z\right)\rho_T/2.
\end{split}
\end{equation}

As pointed out in \cite{cooper}, in the general case where $\eta$ and $\xi$ are
 different from 1, we have seen from the expression $E_{\theta_1,\theta_2}$ that the effective velocity in each component is different. Nevertheless, in the case when $\eta=\xi=1$, it is possible to define an effective velocity shared by both components,
\begin{equation}
    v_{\text{eff}}=\frac{\nabla\Theta}{2}+\frac{\bm{R}S_z}{2(1-S_z^2)}
\end{equation} where $\Theta=\theta_1+\theta_2$ and $\bm{R}=S_y\nabla S_x-S_x\nabla
S_y$. We note the identity
\begin{equation}
    \frac{(\nabla S_z)^2}{(1-S_z^2)}=(\nabla \bm{S})^2-\frac{R^2}{(1-S_z^2)},
\end{equation} where $(\nabla \bm{S})^2=(\nabla S_x)^2+(\nabla S_y)^2+(\nabla S_z)^2$.
Expansion of the square in $E_{\theta_1,\theta_2}$,
 substituting in the $v_{\text{eff}}$ and using the identity
from above reduces the energy to the simple form found in \cite{ktu,ktu3}.
\begin{equation}
\begin{split}
\label{sig2}
    E=&\int\frac{1}{2}(\nabla\sqrt{\rho_T})^2+\frac{\rho_T}{8}(\nabla\bm{S})^2\\
 %&+\frac{\rho_T}{2}\left(\nabla\Theta/2+\bm{R}S_z/2(1-S_z^2)-\bm{\Omega}\times\bm{r}\right)^2\\
&\quad+\frac{\rho_T}{2}\left(v_{\text{eff}}-\bm{\Omega}\times\bm{r}\right)^2+\frac{1}{2}r^2(1-\Omega^2)\rho_T\\
&\qquad+\frac{\rho_T^2}{2}\left(c_0+c_1S_z+c_2S_z^2\right)\quad d^2r,
\end{split}
\end{equation}
where ${c}_0$, ${c}_1$ and ${c}_2$ are equal to $\bar{c}_0$, $\bar{c}_1$ and $\bar{c}_2$ with $\eta$ set equal to unity.
%The chemical potential follows from Eq. (\ref{chemm}) to be $\mu=|\chi_1|^2\mu_1+|\chi_2|^2\mu_2$.

\subsection*{B. A Thomas-Fermi approximation}

The typical Thomas-Fermi (TF) approximation to Eq.'s (\ref{muu1})-(\ref{szeq1}) is
to assume that derivatives in $\rho_T$ and $S_z$ are negligible in
front of the other terms. If we apply the TF approximation we then get
\begin{equation}
    \begin{split}
    \label{muu}
    \mu+\lambda S_z=&\left[(j_1+j_2/\eta)+(j_1-j_2/\eta)S_z\right]r^2\\
&\quad+\rho_T(\bar{c}_0+\bar{c}_1S_z+\bar{c}_2S_z^2),
\end{split}
\end{equation}
and
%The Euler-Lagrange equation satisfying $S_z$ reduces to
\begin{equation}
    \label{szeq}
    \lambda=(j_1-j_2/\eta)r^2+\frac{1}{2}\left({\bar{c}_1}+2\bar{c}_2S_z\right)\rho_T.
%-\frac{\mu_1-\mu_2/\eta}{2}=0.
\end{equation}
The TF energy is then
 \begin{equation}
    \label{sigTF}
\begin{split}
E_{\text{TF}}=&\int \left[(j_1+j_2/\eta)+(j_1-j_2/\eta)S_z\right]r^2\rho_T\\
&\quad+\frac{\rho_T^2}{2}\left(\bar{c}_0+\bar{c}_1S_z+\bar{c}_2S_z^2\right)-\mu\rho_T
-\lambda \rho_T S_z\quad d^2r.
\end{split}
\end{equation}It is important to point out that the reduction of this quadratic form in $\rho_T$ and
$\rho_T S_z$ yields
\begin{equation}
    \label{sigTFred}
\begin{split}
E_{\text{TF}}=&\int \frac{\bar c_2}2 \left(\rho_TS_z+\frac{\bar
c_1}{2\bar c_2}\rho_T+
\frac 1{\bar c_2}(j_1-j_2/\eta)r^2-\frac{\lambda}{\bar c_2}\right )^2\\
&\quad+\frac 12 \left(\bar c_0-\frac {\bar c_1^2}{4\bar c_2}\right)\times\\
&\qquad \left[ \rho_T-\frac 1 {\left(\bar c_0-\bar c_1^2/4\bar
c_2\right)}(\mu -(j_1+j_2/\eta)r^2) \right .\\
 &\quad\quad\left . +\frac 12\frac{\bar c_1}{(\bar c_0\bar c_2- \bar c_1^2/4)}(\lambda
-(j_1+j_2/\eta)r^2)\right ]^2\ d^2r\\
&\qquad\quad+\text{constant terms}.
\end{split}
\end{equation} The existence of a minimum for this quadratic form
implies that $\bar c_2 \geq 0$ and $\bar c_0- \bar c_1^2/(4\bar c_2)
\geq 0$. Since $\bar c_0- \bar c_1^2/(4\bar c_2)
=g_1g_2\Gamma_{12}/(4\eta^2\bar c_2)$, this implies in particular
that $\Gamma_{12}>0$.

Therefore, if $\Gamma_{12}>0$, (which implies $\bar c_2>0$), a
Thomas Fermi approximation can be performed as has previously been considered,
 generally taking an  approximation on the individual component
wave functions $\psi_1$ and $\psi_2$
\cite{ohberg1,tripp,corro,hs,kyt,pb,riboli,jc}. Nevertheless, as we
will see below, if $\bar c_2<0$ and $\Gamma_{12}<0$, then $(\bar c_0-\frac {\bar c_1^2}{4\bar c_2})>0$
 and a TF
approximation can still be performed on $\rho_T$, provided we keep
gradient terms in $S_z$.

 Multiplying (\ref{szeq}) by $S_z$ and subtracting (\ref{muu})
 leads to
 \begin{equation}
    \mu=(j_1+j_2/\eta)r^2+\rho_T\left(\frac{\bar{c}_1}2S_z+\bar{c}_0\right).
\end{equation}
 Simplification by $\rho_T S_z$ with (\ref{szeq}) yields, when
 $\psi_1\times\psi_2\neq0$,
  \begin{equation}
    \label{rhoty}
\rho_T=\frac{a_3+a_4r^2}{g_1g_2\Gamma_{12}},
\end{equation}
and
\begin{equation}
    \label{szform}
S_z=\frac{a_1+a_2r^2}{a_3+a_4r^2},
\end{equation}
where
\begin{subequations}
    \label{as}
\begin{align}
    &a_1=4\eta^2(\lambda\bar{c}_0-\mu\bar{c}_1/2),\\
%    &a_1=\mu_1(g_2+\eta g_{12})-\mu_2(\eta g_1+g_{12}),\\
&a_2=\eta g_1h_2-g_2h_1,\\
%    \begin{split}
%    &a_2=2\gamma_1[(g_1+g_{12})-\eta\xi^2(g_2+g_{12})]\\
%    &\quad-2\gamma_2[(g_1+g_{12})-\eta(g_2+g_{12})]\Omega^2,
%\end{split}\\
&a_3=4\eta^2(\mu\bar{c}_2-\lambda\bar{c}_1/2),\\
%    &a_3=\mu_1(g_2-\eta g_{12})+\mu_2(\eta g_1-g_{12}),\\
&a_4=-(\eta g_1h_2+g_2h_1),
%    \begin{split}
%    &a_4=-2\gamma_1[(g_1-g_{12})+\eta\xi^2(g_2-g_{12})]\\
%    &\quad+2\gamma_2[(g_1-g_{12})+\eta(g_2-g_{12})]\Omega^2,
%\end{split}
\end{align}
\end{subequations}
and where we define
\begin{equation}
    h_k=2\left(j_k-\frac{g_{12}}{g_{3-k}}j_{3-k}\right).
\end{equation}
When only one component is present ($\psi_1\times\psi_2=0$), this
simplifies to
\begin{equation}\label{rhoonly}
    \rho_T=
\begin{cases}
\frac{1}{g_1}\left[\mu+\lambda-2j_1 r^2\right]&\hbox{ if } \psi_{2}=0\\
\frac{\eta}{g_2}\left[\eta(\mu-\lambda)-2j_2r^2\right]&\hbox{ if } \psi_1=0
\end{cases}
\end{equation}
since $S_z=+1$ when $\psi_2=0$ and $S_z=-1$ when $\psi_1=0$.

To begin, we note from Eq. (\ref{szform}) that
$r^2=(a_1-a_3S_z)/(a_4S_z-a_2)$, so that
\begin{equation}\label{rhoSz}
    \rho_T=\frac{a_1a_4-a_2a_3}{g_1g_2\Gamma_{12}(a_4S_z-a_2)}.
\end{equation}
Since the intracomponent coupling strengths are chosen to be
distinct, the support of each component will not necessarily be
equal. In order to
 lead the computations, we have to assume a geometry for the
 components: two disks, a disk and annulus or two half-balls (droplets)
 and a sign for $\Gamma_{12}$.

\subsection*{C. $\Gamma_{12}>0$}

\subsubsection*{1. Both components are disks}

We start when both components are disks.  Without loss of
generality, we can assume that the outer boundary of component-2 (at
$r=R_2$) is larger than that of component-1 (at $r=R_1$). Therefore,
$S_z=-1$ and (\ref{rhoonly}) holds in the annulus, while (\ref{rhoSz}) holds in the coexisting region, which is the inside disk. The integrals from Eq.'s (\ref{nn2}) then
give
\begin{subequations}
    \begin{equation}
    \begin{split}
            \label{szp1}
    \frac{{N}_1}{2\pi}=&\int_0^{R_1}\rho_T\frac{(1+ S_z)}{2}\quad rdr\\
    =&-\frac{(a_1a_4-a_2a_3)^2}{4g_1g_2\Gamma_{12}}\int_{s_0}^{-1}\frac{(1+ S_z)}{(a_4S_z-a_2)^3}\quad dS_z,
\end{split}
\end{equation}
where $s_0=S_z|_{r=0}=a_1/a_3$ and
    \begin{equation}
    \begin{split}
        \label{szp1b}
    \frac{\eta{N}_2}{2\pi}=&\int_0^{R_1}\rho_T\frac{(1- S_z)}{2}\quad rdr+\int_{R_1}^{R_2}\rho_T\quad rdr\\
    =&-\frac{(a_1a_4-a_2a_3)^2}{4g_1g_2\Gamma_{12}}\int_{s_0}^{-1}\frac{(1- S_z)}{(a_4S_z-a_2)^3}\quad dS_z\\
    &\quad-\frac{g_2}{4\eta j_2}\int_{\rho_d}^0\rho_T\quad d\rho_T,
\end{split}
\end{equation}
\end{subequations}
with
    \begin{equation}
    \begin{split}
    \rho_d=&\frac{a_3+a_4r^2}{g_1g_2\Gamma_{12}}\Big|_{r=R_1}\\
    =&\frac{a_2a_3-a_1a_4}{g_1g_2\Gamma_{12}(a_2+a_4)},
\end{split}
\end{equation}
since $S_z=-1$ at $r=R_1$.

Completion of these integrals and noting that $\{a_2+a_4,a_2-a_4\}=\{-2g_2h_1,2\eta g_1h_2\}$ gives
\begin{subequations}
    \label{all}
\begin{equation}
    \label{szp2}
    a_3^2=
%   \begin{cases}
\frac{8N_1g_1g_2^2\Gamma_{12}h_1}{\pi(s_0+1)^2}\\%\qquad &k=1\\
%\frac{8N_2\eta^2g_1^2g_2\Gamma_{12}h_2}{\pi(s_0-1)^2}\qquad &k=2
%\end{cases}
\end{equation}
which must necessarily be positive, i.e. $h_1\Gamma_{12}>0$, and
\begin{equation}
    \label{qq1}
    \begin{split}
N_1(a_2-s_0a_4)^2=&2\eta^2N_2g_1g_2j_2h_1\Gamma_{12}(1+s_0)^2\\
&\quad+2N_1(1+s_0)\eta j_2g_1\Gamma_{12}\times\\
&\qquad[\eta g_1h_2(1+s_0)+2g_2h_1(s_0-1)]
\end{split}
\end{equation}
\end{subequations}
where in Eq. (\ref{qq1}) the expression for $a_3$ from Eq.
(\ref{szp2}) has been substituted. This equation can always be
solved in terms of $s_0$ when $h_1 \Gamma_{12}$ is positive since
the discriminant is equal to
$8N_1\eta^2g_2^2g_1j_2h_1\Gamma_{12}(N_1g_{12}+N_2 g_2)$.
 We find that $$1+s_0=\frac{2N_1g_2}{N_1g_2-\eta N_1  g_{12}+\sqrt{2N_1\eta^2g_1j_2\Gamma_{12}(N_1g_{12}+N_2 g_2)/h_1}}.$$

A similar calculation can be completed if the outer boundary of
component-2 is larger than that of component-1. This then gives that
${h_2\Gamma_{12}>0}$ (although expressions (\ref{all}) change
slightly).

 If an annulus develops  in
 component $3-k$, it means that $s_0=1$ ($s_0=-1$) for $k=1$ ($k=2$).
  Inputting this choice into Eq. (\ref{qq1})  gives
\begin{equation}
    \begin{split}
    \label{spatsep}
&\bar{g}_{12}=\frac{N_kg_kj_{3-k}}{2(N_1j_1+N_2j_2)}\\
&\quad+\frac{1}{2}\left(\left[\frac{N_kg_kj_{3-k}}{N_1j_1+N_2j_2}\right]^2+\frac{4N_{3-k}g_1g_2j_{3-k}}{N_1j_1+N_2j_2}\right)^{1/2}
    \end{split}
\end{equation}
as the critical $g_{12}$ at which an annulus forms in
component-$\{3-k\}$. In the case of Fig. \ref{sch_3}, the curve (\ref{spatsep}) has been
plotted in dashed lines and is close to the numerical curve.
 Notice that if $\xi^2=1$ (equal trapping
frequencies for both components), then Eq. (\ref{spatsep}) becomes
independent of $\Omega$, as in the phase diagrams of
Fig. \ref{sch_2}, where it yields
 $\Gamma_{12}=0.008$.

Let us point out that before the transition to the disk plus annulus takes place, there is
 a subregion of region 1, where there are 2 disks, but in one component the wave function has a local minimum
  at the origin. For instance, in the case of Fig. \ref{sch_2}, it corresponds to $\Gamma_1$ changing sign.

As a conclusion, in order for the ground state to be composed of two
disks, assuming that component-$k$ is the component with smaller
support, we need that $h_k>0$, $\Gamma_{12}>0$ and
$g_{12}<\bar{g}_{12}$. These three conditions can be summarised as
\begin{subequations}
\begin{align}
\label{con1a}
%\begin{cases}
g_{12}<\min\left(\frac{j_k}{j_{3-k}}g_{3-k},\frac{j_{3-k}}{j_{k}}g_{k},\bar{g}_{12},\sqrt{g_1g_2}\right)
\hbox{ if } &h_1,\ h_2>0,\\
\frac{j_{3-k}}{j_{k}}g_{k}<
g_{12}<\min\left(\frac{j_k}{j_{3-k}}g_{3-k},\bar{g}_{12},\sqrt{g_1g_2}\right)\hbox{
if } &h_{3-k}<0.
%\end{cases}
\label{con1b}
\end{align}
\end{subequations}
%\begin{figure}
%\begin{center}
%\includegraphics[scale=0.5]{phase_summary.eps}\\
%\end{center}
%\caption{An $\Omega-\Gamma_{12}$ phase diagram depicting the possible regions associated with a Thomas-Fermi analysis on both wave functions. The parameters are $g_1=0.01$, $g_2=0.003$, $\eta=0.6$, $\xi=1.55$, $N_1=N_2=10^6$. The regions are: (I) two disks with $h_1>0$, $h_2>0$ satisfying Eq. (\ref{con1a}). The boundary at which $R_1=R_2$ is denoted by the dash-dot line ($h_1N_2g_2=h_2N_1g_1$) and the boundary when $g_k=g_{3-k}\Lambda_k$ is denoted by the dash line (Eq. (\ref{toget})); (II) two disks with $h_1<0$ and $h_2>0$ satisfying Eq. (\ref{con1b}) with $k=2$ so that component-1 has a dip at the centre; (III) two disks with $h_1>0$ and $h_2<0$ satisfying Eq. (\ref{con1b}) with $k=1$ so that component-2 has a dip at the centre; (IV) an annulus in component-1 and a disk in component-2 with $h_1<0$ and $h_2>0$ satisfying Eq. (\ref{an}) with $k=2$ and (V) an annulus in component-2 and a disk in component-1 with $h_1>0$ and $h_2<0$ satisfying Eq. (\ref{an}) with $k=1$. The boundary between region (I) and (II) is given by $h_1=0$, between region (I) and (III) by $h_2=0$, between region (II) and (IV) by Eq. (\ref{spatsep}) with $k=1$ and between region (III) and (V) by Eq. (\ref{spatsep}) with $k=2$. The value of $\Omega^{\text{lim}}=0.78$ (dotted line) and all the boundaries meet at the location $(\Gamma_{12}^*,\Omega^*)=(0,0.44)$.}
%\label{phase1}
%\end{figure}

\subsubsection*{2. A disk and an annular component}

In this case, we can assume a disk in component-1 and an annulus in component-2; there are three regions:
 an inner disk where only component-1 is present and (\ref{rhoonly}) holds, an outer
  annulus where only component-2 is present and (\ref{rhoonly}) holds and an inner
  annulus where both components coexist and (\ref{rhoSz}) holds.

In order to use the TF approximation when one component is annular,
 computations similar to
(\ref{szp2})-(\ref{qq1}) lead to $h_1h_2<0$ and
$g_{12}>\bar{g}_{12}$. This can be summarized (for an annulus in
component-$\{3-k\}$) as
\begin{equation}
    \label{an}
    \max\left(\frac{j_{3-k}}{j_{k}}g_{k},\bar{g}_{12}\right)< g_{12}<\min\left(\frac{j_k}{j_{3-k}}g_{3-k},\sqrt{g_1g_2}\right),
\end{equation}
with $h_{3-k}<0$.
Equation (\ref{an}) also places the restriction that
\begin{align}
    g_k<&g_{3-k}\left(\frac{j_k}{j_{3-k}}\right)^2\nonumber\\
    =&g_{3-k}\Lambda_k,
    \label{toget}
\end{align}
where $\Lambda_k=({j_k}/{j_{3-k}})^2$. Note that $\Lambda_1\Lambda_2=1$ such that an annulus develops in component-2 (-1) if $g_2 >$ ($<$) $g_1\Lambda_2$.

\subsubsection*{3. Orders of Intracomponent Strengths and Special Cases}

 The effect that changing the order of the intracomponent
strengths and particle numbers has on $\bar{g}_{12}$, and thus on
the phase diagrams, is now investigated. There are two cases to
consider, depending on the relative orders of the particle numbers.
Drawing aid from experimental values, it is always expected that
$\min \{N_1, N_2\}\gg \max \{ g_1,\ g_2\}$. Throughout it will be
assumed that $g_2>\Lambda_2 g_1$, so the annulus develops in
component-2 and that $j_1$ and $j_2$ are of order unity.
In the first case when the particle number of  component-1 is much
greater than the particle number of component-2 ($N_1\gg N_2$), it
follows that  $\bar{g}_{12}\sim g_1j_2/j_1$.
%Then, in the limit $g_{12}\rightarrow\bar{g}_{12}$,
%\begin{equation}
%   \Gamma_{12}\sim1-\frac{g_1(2\gamma_1-\Omega^2)^2}{g_2(2\gamma_1\gamma_2-\Omega^2)^2}.
%\end{equation}
The boundary between  region 1 and region 3 is then directly
dependent on the value of the ratio $j_2g_1/[j_1g_2]$. Notice that
if $g_2\gg g_1$, $\Gamma_{12}$ evaluated at $g_{12}=\bar{g}_{12}$
tends to unity and as such an annulus would always be present in
component-2, whatever the value of $g_{12}$.
%Furthermore the radii (Eq.'s (\ref{eq:r11v})-(\ref{eq:r21v})), in the limit $\Gamma_{12}\rightarrow0$ and $\mathcal{O}(g_2)\gg\mathcal{O}(g_1)$, all reduce to
%\begin{equation}
%   R_1,R_{2_-},R_{2_+}\sim\left(\frac{4N_1g_1}{\pi(2\gamma_1\gamma_2-\Omega^2)}\right)^{1/4}.
%\end{equation}
%Therefore in the case where $\mathcal{O}(N_1)\gg \mathcal{O}(N_2)$ and $\mathcal{O}(g_2)\gg \mathcal{O}(g_1)$ (providing $\mathcal{O}(N_2)\gg \mathcal{O}(g_2)$), component-1 is large and disk-shaped, whereas component-2 is a thin annulus that is localised around the radius of component-1.
%
%\subsubsection*{2. $\mathcal{O}(N_1)\ll \mathcal{O}(N_2)$}

Conversely in the second case when the particle  number of
component-1 is much smaller than the particle number of component-2
($N_1\ll N_2$), it follows that  $\bar{g}_{12}\sim\sqrt{g_1g_2}$
which implies that the annulus will only develop near
$\Gamma_{12}=0$.

%\begin{figure}
%\begin{center}
%\includegraphics[scale=0.5]{tfprofiles1.eps}
%\end{center}
%\caption{Total density profiles (divided by $10^4$) of the two components calculated numerically (solid lines) and analytically from the TF approximation (dashed lines). All plots are at $\Omega=0$ along $y=0$ and are for the case where (a) $g_1=0.0078$, $g_2=0.0083$, $N_1=N_2=10^5$ and $\eta=\xi=1$ (case ES1)  and (b) $g_1=0.0067$, $g_2=0.0063$, $N_1=N_2=10^4$, $\eta=0.48$ and $\eta\xi^2=1$ (case ES2). The value of $\Gamma_{12}$ is $0.5$. Distance is measured in units of $r_0$ and density in
%units of $r_0^{-2}$.}\label{tfprofiles}
%\end{figure}

%\subsubsection*{4. Special cases}

We can also look at some special cases - there are four that can be considered:

{\it Case (i)}. $\Lambda_k g_{3-k}=g_k$. When $\Lambda_2$
(equivalently $\Lambda_1$) is such that this equality is made, the
two components are both disks and no annulus develops.

{\it Case (ii)}. $\eta=\xi=1$. Then $\Lambda_k=1$ %and $S_z$ reduces to
%\begin{equation}
%   S_z=\frac{\mu_1(g_2+g_{12})-\mu_2(g_1+g_{12})+c_1(1-\Omega^2)r^2}{\mu_1(g_2-g_{12})-\mu_2(g_1-g_{12})-2c_2(1-\Omega^2)r^2}
%\end{equation}
and $R_2\gtrless R_1\iff N_2g_2\Gamma_1\gtrless N_1g_1\Gamma_1$ when
there are two disks and the annulus develops in the component which
has the larger interaction strength.

{\it Case (iii)}. $\eta=\xi=1$ and $g_1=g_2\equiv g$. When the
intracomponent coupling strengths are equal, $\bar{g}_{12}=g$ and
there are always two disks with $R_2\gtrless R_1\iff N_2\gtrless
N_1$.

{\it Case (iv)}. $\eta=\xi=1$, $g_1=g_2\equiv g$ and $N_1=N_2\equiv
N$. When the particle numbers and intracomponent coupling strengths
are equal, $\bar{g}_{12}=g$  and there are always two disks with
$R_2=R_1$. A detailed analysis of this case, explored numerically in
a phase diagram for all $\Gamma_{12}$ and analytically in the TF
limit, was considered by \cite{ktu1}.

\subsubsection*{4. Justification of the Thomas-Fermi approximation}

  The gradient terms in
(\ref{muu1})-(\ref{szeq1}) can be neglected if $d_{hl}$,
 the characteristic length of variation of  $\rho_T$ and $S_z$ is much smaller than
$d_c$, the characteristic size of the condensates. We have that
$1/d_{hl}^2$ is of order of  $\mu$ and $\lambda$, which are of order
$\sqrt{N_kg_k}$, $\sqrt{N_kg_{3-k}}$. Hence
 $d_{hl}$ is bounded above by the maximum
 of $(N_kg_k)^{-1/4}$, $(N_kg_{3-k})^{-1/4}$. From the expression of $a_3$,
the characteristic size of the condensate is of order the minimum of
$(g_kN_k \Gamma_{12})^{1/4}$. Therefore, the Thomas Fermi approximation holds
 if $N_kg_k\sqrt{\Gamma_{12}}$, $N_kg_{3-k}\sqrt{\Gamma_{12}}$ are large. This requires
 the usual Thomas Fermi criterion that $N_kg_k$, $N_kg_{3-k}$ are large, but
 breaks down if $\Gamma_{12}$ is too small.

For $\Gamma_{12}=0$,
 the
 equations lead to spatial separation: either the radii of the disks tend to 0
 in the case of two disks or the outer radius of the inner disk tends
 to the inner radius of the annulus in the case of disk plus annulus
 so that $\psi_1\times\psi_2=0$ everywhere. Another analysis
 has to be carried out to understand the region of coexistence,
  which is of small size and has strong gradients.

\subsection*{D. $\Gamma_{12}<0$, beyond the TF approximation}

For negative $\Gamma_{12}$, and if $N_kg_k$ are large, the  Thomas Fermi
  approximation can be extended provided some model takes into account the small region where
the condensates coexist.

Indeed, if we go back to (\ref{sigTFred}), and have both $\bar c_2<0$ and $\Gamma_{12}<0$,
 then the coefficient in front of the second square is positive and the optimal situation
 is to have the square equal to 0, which leads to the inverted parabola (\ref{rhoty}). On the other
 hand, the coefficient in front of the first square is negative, and the ground state involves
  derivatives  in $S_z$ to compensate it. Under the assumption
   that the boundary layer where $S_z$ varies is small,
 we are going to derive a TF model with jump  for $\rho_T$. We will analyze it
   for the different geometries (disk plus annulus, droplets and vortex sheets)
    and show that it provides  information consistent  with the numerics.

% On the other hand, for small $\Gamma_{12}$,
%   a model has to be introduced to take into account the boundary
%   layer.

\subsubsection*{1. Disk Plus Annulus}

%emin=3022238.58771168
% A typical example, for $\Gamma_{12}=-0.3$ with the parameters $g_1=0.0078$, $g_2=0.0083$ and $N_1=N_2=10^5$ where $\eta=\xi=1$ (set ES1) gives $r_s=?$ and $k=?$. A plot of the density profile with these values %($\mu_1=?$ and $\mu_2=?$)
% is given in Fig \ref{sig1}(a), which shows that the numerical and analytical density profiles agree well. %emin=?

 Assuming that the
boundary layer is present only at some $r=r_s$, then $S_z=+1$ in the region in which component-2 is zero
($r\le r_s^-$) and $S_z=-1$ in the region in which component-1 is
zero ($r\ge r_s^+$). The
transition from $S_z=+1$ to $S_z=-1$ is not smooth, therefore
creating the jump in density.
%\begin{equation}
%   \begin{split}
%\label{match}
%   E=&\int_\mathcal{D}\frac{1}{2}\left(\gamma_1\left(1-S_z+\gamma_2(1+S_z)\right)-\Omega^2\right)r^2\r%ho_T\\
%   &+\frac{\rho_T^2}{2}\left(c_0+c_1S_z+c_2S_z^2\right)\quad d^2r,
%\end{split}
%\end{equation}
%with
%\begin{eqnarray}
%S_z&=&+1 \quad\text{for}\quad r\le r_0^-,\nonumber\\
%S_z&=&-1 \quad\text{for}\quad r\ge r_0^+.
%\end{eqnarray}
% \begin{figure}
% \begin{center}
% \includegraphics[scale=0.5]{sig_1.eps}
% \end{center}
% \caption{Total density profiles (divided by $10^4$) for $\Gamma_{12}=-0.05$ with $g_1=0.0078$, $g_2=0.0083$ and $N_1=N_2=10^5$, $\eta=\xi=1$ and $\Omega=0$ (set ES1) obtained numerically (solid line) and analytically (a) by taking an asymptotical form for $\bm{S}$ (dashed line) and (b) for a density jump over the boundary layer (dotted lines). The inset shows the discontinuity of density at $r_s=3.62$. Distance is measured in units of $r_0$ and density in
% units of $r_0^{-2}$.}\label{sig1}
% \end{figure}
\begin{figure}
\begin{center}
\includegraphics[scale=0.5]{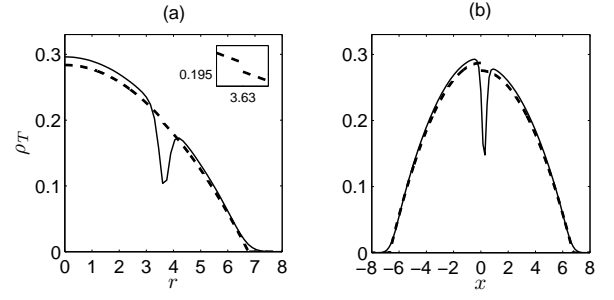}
\end{center}
\caption{Total density profiles (divided by $10^4$) obtained numerically (solid lines) and analytically (dashed lines) for two spatial separation cases with $g_1=0.0078$, $g_2=0.0083$,  $N_1=N_2=10^5$, $\eta=\xi=1$ (set ES1) and $\Gamma_{12}=-10$ and $\Omega=0$: (a) annulus plus disk,  analytical estimate coming from (\ref{eann}) and (b) droplet, analytical estimate coming from (\ref{edrop}). The inset in (a) shows the discontinuity of density at $r_s=3.63$.  Distance is measured in units of $r_0$ and density in units of $r_0^{-2}$.}\label{sig1}
\end{figure}
Therefore, we are lead to minimise the integral
\begin{equation}
    \label{eann}
    \begin{split}
I&=\int_{B_{r_s}}2j_1r^2\rho_T+\frac{g_1}{2}\rho_T^2 \quad d^2r\\
&\quad+\int_{B_{R\backslash r_s}}2\frac{j_2}{\eta}r^2\rho_T+\frac{g_2}{2\eta^2}\rho_T^2\quad  d^2r
\end{split}
\end{equation}
with respect to $r_s$. Here $B_{r_s}$ is a ball of radius $r_s$ and $B_{R\backslash r_s}$ is a torus with outer boundary at $r=R$ and inner boundary at $r=r_s$. Thus
\begin{equation}
    \begin{split}
&2\pi r_s\left[2j_1r^2\rho_T+\frac{g_1}{2}\rho_T^2\right]\Bigg|_{r=r_s^-}\\
&\qquad-2\pi r_s\left[2\frac{j_2}{\eta}r^2\rho_T+\frac{g_2}{2\eta^2}\rho_T^2\right]\Bigg|_{r=r_s^+}=0,
\label{jumpp}
\end{split}
\end{equation}
which implies that
\begin{subequations}
    \label{rhopm}
\begin{align}
\rho^-=&\frac{1}{g_1}\left[\mu_1-2j_1r^2\right]\quad r\in B_{r_s}\\
\rho^+=&\frac{\eta}{g_2}\left[\mu_2-2j_2r^2\right]\quad r\in B_{R\backslash r_s}.
\end{align}
\end{subequations}
Then using the normalisation conditions $\int\rho^-d^2r=N_1$ and $\int\rho^+d^2r=\eta N_2$ we get an outer radius
\begin{subequations}
    \begin{equation}
        R=\left(r_s^2+\sqrt{\frac{g_2N_2}{\pi j_2}}\right)^{1/2},
    \end{equation}
and chemical potentials
\begin{align}
\mu_1=&\frac{N_1g_1}{\pi r_s^2}+j_1r_s^2,\\
\mu_2=&2j_2\left(r_s^2+\sqrt{\frac{g_2N_2}{\pi j_2}}\right).
\end{align}
\end{subequations}
%where $\rho_0^-=\rho^-\big|_{r=r_0^-}$ and $\rho_0^+=\rho^+\big|_{r=r_0^+}$.
%\begin{equation}
%\frac{g_1N_1}{\pi}=j_1r_0^4+\rho_{0}^-g_1r_0^2,
%\end{equation}
%Similarly, by $\int\rho^+d^2r=N_2$,
%\begin{subequations}
%\begin{eqnarray}
%\frac{g_2N_2}{\pi}&=&\frac{(1+\eta)}{8\eta}\left(\frac{4}{(1+\xi)^2}-\Omega^2\right)(R^2-r_0^2)^2,\\
%\frac{g_2N_2}{\pi}&=&-\frac{(1+\eta)}{8\eta}\left(\frac{4}{(1+\xi)^2}-\Omega^2\right)(R^2-r_0^2)^2\nonumber\\
%&&\quad-(r_0^2-R^2)\rho_{0}^+g_2,
%\end{eqnarray}
%\end{subequations}
%where . Now, these two equations imply that
%\rho_0^-&=&\frac{N_1}{\pi r_0^2}-\frac{j_1r_0^2}{g_1},\\
%\rho_0^+&=&2\sqrt{\frac{N_2j_2}{\pi g_2}},\\
It remains to find $r_s$. But, from (\ref{jumpp}), it follows that
\begin{equation}
r_s^2=\frac{\frac{4N_2j_2}{\pi}-g_1\left(\frac{N_1}{\pi r_s^2}-\frac{j_1r_s^2}{g_1}\right)^2}{4\left[j_1\left(\frac{N_1}{\pi r_s^2}-\frac{j_1 r_s^2}{g_1}\right)-2j_2\sqrt{\frac{N_2j_2}{\pi g_2}}\right]}.
\end{equation}
The above leaves a quartic equation in $r_s^2$, the solution of which can be found numerically.

We can then complete the energy calculation to give
\begin{equation}
        \label{enda}
I=\frac{g_1N_1^2}{2\pi r_s^2}+(N_1j_1+2N_2j_2)r_s^2-\frac{\pi j_1^2}{6g_1}r_s^6+\frac 43\left(\frac{N_2^3g_2j_2}{\pi}\right)^{1/2}.
\end{equation}
%\begin{eqnarray}
%&&\frac{3(1-\Omega^2)^2}{16}r_0^8+(1-\Omega^2)\sqrt{\frac{N_2g_1^2(1-\Omega^2)}{\pi g_2}}r_0^6\nonumber\\
%&&\quad-\frac{g_1(1-\Omega^2)}{2\pi}(N_1-2N_2)r_0^4-\frac{N_1^2g_1^2}{\pi^2}=0.
%\end{eqnarray}
A check plot of the density profile with the parameters $g_1=0.0078$, $g_2=0.0083$ and $N_1=N_2=10^5$ where $\eta=\xi=1$
%is given in Fig \ref{sig1}(b),
agrees well with this model
 where the boundary is calculated to be at $r_s=3.62$ (see Fig.\ref{sig1}(a)).

\subsubsection*{2. Droplets}

This formalism can also be extended to study the droplet case. As before, assuming a thin boundary layer, one can allow for a jump in density. We thus need to minimise the Thomas-Fermi energy for two droplets described by $B_{1}$ and $B_{2}$ that exist in the regions $0\le\theta\le\alpha$ and $\alpha\le\theta\le2\pi$ respectively. The energy is then
\begin{equation}
    \label{edrop}
    \begin{split}
I&=\alpha\int_{0}^{R_1}2j_1r^2\rho_T+\frac{g_1}{2}\rho_T^2 \quad rdr\\
&\quad+(2\pi-\alpha)\int_{0}^{R_2}2\frac{j_2}{\eta}r^2\rho_T+\frac{g_2}{2\eta^2}\rho_T^2\quad  rdr.
\end{split}
\end{equation}
The expressions for the density in each domain (\ref{rhopm}) and the normalisation conditions allow completion of this integral to give
\begin{equation}
I=\frac{4\sqrt{2}}{3}\left[\left(\frac{N_1^3g_1j_1}{\alpha}\right)^{1/2}+\left(\frac{N_2^3g_2j_2}{(2\pi-\alpha)}\right)^{1/2}\right].
\end{equation}

It remains to find the optimum $\alpha$. This is achieved through the condition $dI/d\alpha=0$ which gives
\begin{equation}
    \begin{split}
\frac{(N_1^3g_1j_1)^{1/2}}{\alpha^{3/2}}&=\frac{(N_2^3g_2j_2)^{1/2}}{(2\pi-\alpha)^{3/2}}\\
\Rightarrow \alpha&=2\pi\frac{\bar{N}(\bar{g}\bar{j})^{1/3}}{(1+\bar{N}(\bar{g}\bar{j})^{1/3})}
\end{split}
\end{equation}
where we have set $\bar{N}=N_1/N_2$, $\bar{g}=g_1/g_2$ and $\bar{j}=j_1/j_2$. As expected, equality of the $N_k$, $g_k$ and setting $\eta=\xi=1$ ($\Rightarrow j_1=j_2$) gives $\alpha=\pi$ and the condensate is then composed of two half-balls. Otherwise, a curvature is present. A check plot of the density profile with the parameters $g_1=0.0078$, $g_2=0.0083$ and $N_1=N_2=10^5$ where $\eta=\xi=1$ is given in Fig \ref{sig1}(b).

Finally we can note the energy for the droplets is
\begin{equation}
	\label{endd}
    \begin{split}
I&=\frac{4}{3\sqrt{\pi}}\left(1+\bar{N}(\bar{g}\bar{j})^{1/3}\right)^{-1/2}\times\\
&\quad\left[(N_1^3g_1j_1)^{1/2}+(N_1N_2^2(g_1g_2^2j_1j_2^2)^{1/3})^{1/2}\right].
\end{split}
\end{equation}
This energy can be compared to the energy of the disk plus annulus (\ref{enda}) to determine which is the optimum geometry.
Indeed, in the numerical cases studied before, the droplets are preferred states for small $\Omega$.
 %A plot of the density profile with the parameters $g_1=0.0078$, $g_2=0.0083$ and $N_1=N_2=10^5$ where $\eta=\xi=1$ is given in Fig \ref{sig1}(c).

\subsubsection*{3. Regions of Vortex Sheets}

In the case of vortex sheets,
 we can assume that the global profile of the total density is TF-like, obeying Eq. (\ref{rhoty}) in the bulk of the condensate. By working with the total density, we do not require any information on the vortex sheets themselves (and consequently $S_z$).

We thus take the form of $\rho_T$ from Eq. (\ref{rhoty}) from which we note that the outer boundary at $r=R$ satisfies $R=\sqrt{-a_3/a_4}$ and that completion of the normalisation condition ($\ref{nn1a}$) gives
\begin{equation}
a_3=\sqrt{-\frac{2(N_1+\eta N_2)g_1g_2\Gamma_{12} a_4}{\pi}}.
\end{equation}
This expression evidently requires $a_4\Gamma_{12}<0$. We however expect the vortex sheets to be present only in the $\Gamma_{12}<0$ domain, thus we can be more specific on the condition, namely that $a_4>0$, or
\begin{equation}
    g_{12}>\frac{\eta g_1j_2+g_2j_1}{(\eta j_1+j_2)}.
\end{equation} We point out that this critical number corresponds to $\Gamma_{12}=0$ in the cases studied above.
With $a_3$ as above and $a_4$ given by the parameters of the system, the density profile is then fully accessible.

% \subsubsection*{4.Small Boundary layer}
% If $\Gamma_{12}$ is too small, one has to take into account the variations in $S_z$ in the boundary
% layer. An expansion of (\ref{szeq}) leads to
% \begin{equation}
%     \label{ansatz}
% S_z=-\tanh[k(r-r_s)]
% \end{equation}
% for some variational parameter $k$ and for $r_s$ to be a location in the boundary layer where $S_z=0$.
% We assume that the annulus is present in component-2. Since $|\bm{S}|^2=1$ and $S_y=0$, this implies that $S_x=\text{sech}[k(r-r_s)]$.
% A typical example, for $\Gamma_{12}=-0.05$ with the parameters $g_1=0.0078$, $g_2=0.0083$ and $N_1=N_2=10^5$ where $\eta=\xi=1$ (set ES1) gives $r_s=3.61$ and $k=0.77$. A plot of the density profile with these values %($\mu_1=22.43$ and $\mu_2=22.86$)
% is given in Fig \ref{sig1}(a), which shows that the numerical and analytical density profiles agree well.

\subsection*{E. Analysis of defects}

The advantage of the Thomas Fermi analysis of $\rho_T$ in the nonlinear sigma
model is that it allows for analysis of
 defects as a perturbation calculation when $N_k g_k$, $N_k g_{3-k}$ and $N_k g_{12}$ are large,
  in the spirit of what has been done for a single condensate in \cite{ad,ar,cd,wei}
  or for two condensates in \cite{wei2}.
   We do not need
   to analyze specifically the peaks because they are taken into account in the $S_z$ formulation.
   This has
been attacked in a more complicated fashion in the case of a single coreless vortex in \cite{ktu}.
   We will develop our ideas  in a later work but let us point out that the starting point
   is the energy (\ref{sig}) and if we call $p_i$ and $q_i$
    the location of the vortex in each component, the main terms coming from the vortex contributions lead to:
   \begin{itemize}\item Each vortex core of each component providing a kinetic energy term
    proportional to $\rho_T(0) \log d_{hl}/(2(1+\eta))$ where $d_{hl}$ is the healing length, the characteristic
     size of the vortex core, here of order $1/\sqrt {N_kg_k}$.
     \item the rotational term providing a term in $-c\Omega (\rho_T(0))^2$, where $c$
        is a numerical constant. The balance of these two terms allows us to compute the critical velocity
         for the nucleation of the first vortex.
         \item the kinetic energy then yielding a term in $-\log |p_i-p_j|-\log|q_i-q_j|$.
         \item the rotation term yielding a term in $\Omega (|p_i|^2+|q_i|^2)$.
     \item the interaction term providing  a term $\rho_T^2 S_z^2$ for the perturbation of
      $S_z$ close to a coreless vortex. The ansatz can be made in several ways
       leading to an interaction term in $e^{-|p_i-q_i|^2}$.
           \end{itemize}
  This leads to a point energy of the type
  $$\sum_i a (|p_i|^2+|q_i|^2+be^{-|p_i-q_i|^2} )-\sum_{i,j}(\log|p_i-p_j|+\log|q_i-q_j|)$$ where
   $a$ and $b$ are related to the parameters of the problem.
  The ground state of such a point energy leads to a square lattice for a sufficient number of points
   and for some range of $a$ and $b$.

\section*{IV. Phase Diagram Under Conserved Total Particle Number}

 In the experiment of Hall
et al \cite{hall} a single component BEC of $^{87}$Rb in the
$|1$,$-1\rangle$ state was initially created, before a transfer of
any desired fraction of the atoms from this $|1$,$-1\rangle$ state
to the $|2$,$1\rangle$ state created the two-component BEC. Thus the
ratio of particle numbers $N_1/N_2$ is controllable experimentally,
with the constraint that $N_1+N_2$ is constant (in the case of the
experiment of \cite{hall}, $N_1+N_2=5\times10^5$). As such,
experimentally, it is possible to keep the individual particle
numbers constant (as in the normalisation condition (\ref{norm})) or
to keep the total particle number $N_1+N_2$ constant, allowing $N_1$
and $N_2$ to vary (as in (\ref{norm12})). We produce an
$\Omega-\Gamma_{12}$ phase diagram for the parameters $g_1=0.003$,
$g_2=0.006$ and $\eta=\xi=1$ using
the normalisation condition given by (\ref{norm12}) with $N_1+N_2=2.1\times 10^5$.
Note that if $g_1=g_2$, $m_1=m_2$ and
$\omega_1=\omega_2$ ($\eta=\xi=1$), then the phase diagram would be
identical to that of Fig. \ref{sch_1} (i.e. the normalisation
condition in this case would not be important). The $\Omega-\Gamma_{12}$ phase diagram is presented in Fig. \ref{sketchv}. There are three distinct regions (determined by the geometry of the ground state):

\begin{figure}
\begin{center}
%    \hskip -1cm
\includegraphics[scale=0.4]{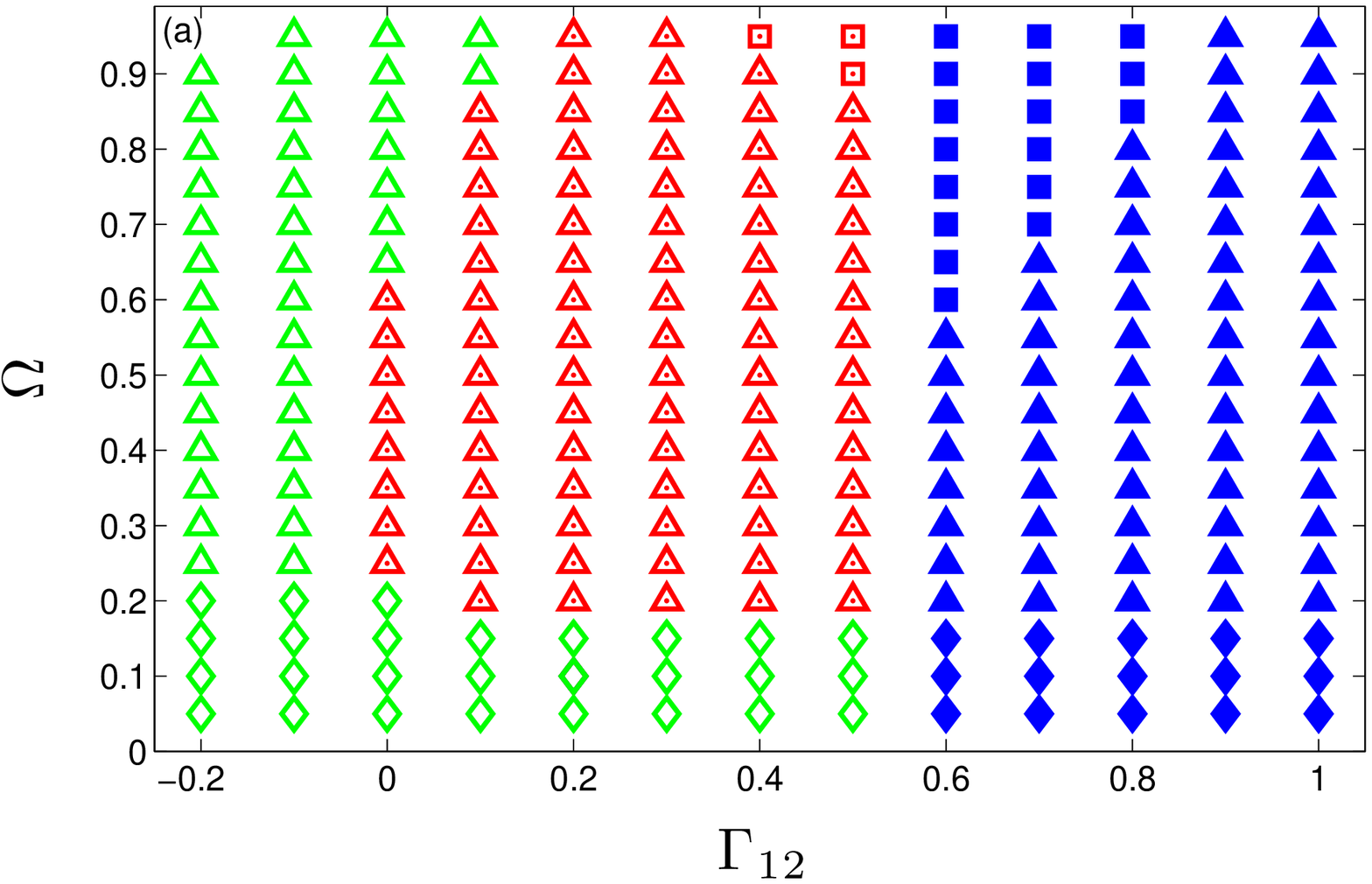}\\
\includegraphics[scale=0.4]{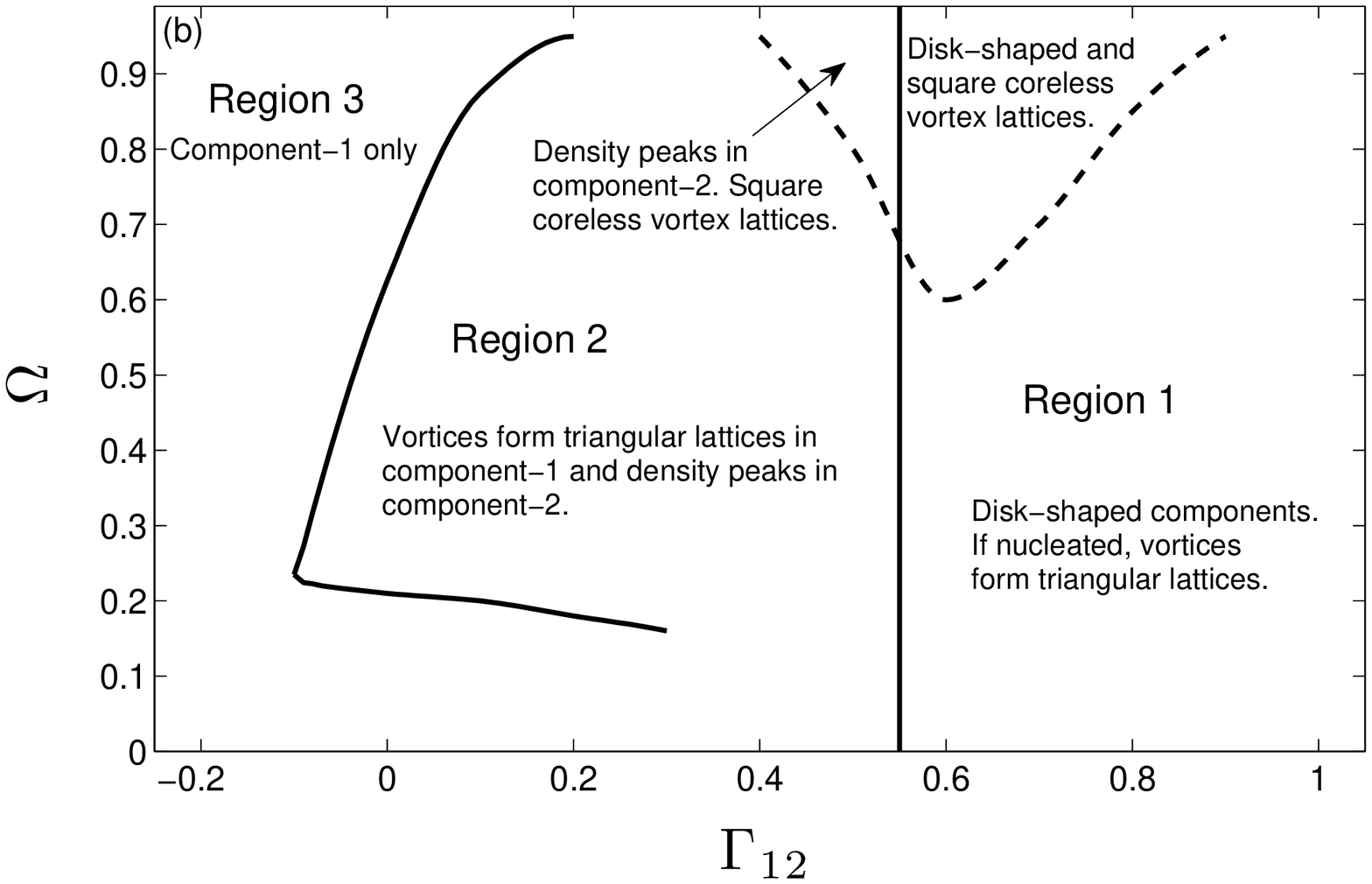}
\end{center}
\caption{(Color online) $\Omega-\Gamma_{12}$ phase diagram for parameters $g_1=0.003$ and $g_2=0.006$
%, $N_1=10^5$, $N_2=1.1\times10^5$
with $\eta=\xi=1$ using the normalisation condition given by (\ref{norm12}). (a) Numerical simulations where triangles indicate that the lattice in both components is triangular, squares that the lattice in both components is square and diamonds where no vortices have been nucleated. Filled triangles, squares and diamonds are where the two components are disk-shaped and coexist, empty triangles, squares and diamonds are where only component-1 exists; those with a dot in the centre represent the appearance of coreless vortices in component-2 and those without a dot in the centre represent the complete disappearance of component-2. (b) A schematic representation of the numerical simulations.
The solid lines indicate the boundary between different identified regions (determined by the geometry of the ground state) and the dashed lines the boundary between triangular and square lattices. The unit of rotation is $\tilde{\omega}$.}\label{sketchv}
\end{figure}

\begin{figure}
\begin{center}
\includegraphics[scale=0.5]{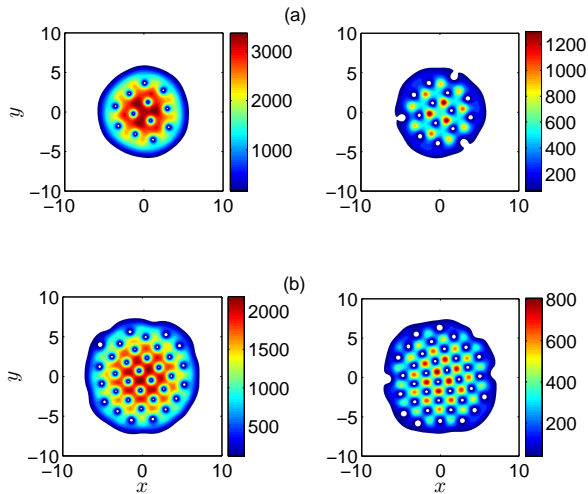}
\end{center}
\caption{(Color online) A series of density plots for component-1 (left column) and component-2 (right column). The parameters are $g_1=0.003$ and $g_2=0.006$ with $\eta=\xi=1$ and $\Gamma_{12}=0.7$ (which gives $g_{12}=0.0023$) and normalisation taken over the total density (Eq. (\ref{norm12})). The angular velocity of rotation is $\Omega$ and it takes the values (a) $0.65$ and (b) $0.9$. There is a triangular lattice in (a) and a square lattice in (b). At these parameters the components are in region 1 of the phase diagram Fig. \ref{sketchv}. Distance is measured in units of $r_0$ and density in
units of $r_0^{-2}$.}
\label{7posv}
\end{figure}

{\it Region 1}. In the first region, both components are disk-shaped. As before, the coreless vortices can either form a triangular or a square lattice depending on the values of $\Gamma_{12}$ and $\Omega$. Figure \ref{7posv} shows this case for $\Gamma_{12}=0.7$ and $\Omega=0.65$ (where a triangular lattice is present) and $\Omega=0.9$ (where a square lattice is present). In region 1, both components have the same radii.

\begin{figure}
\begin{center}
\includegraphics[scale=0.5]{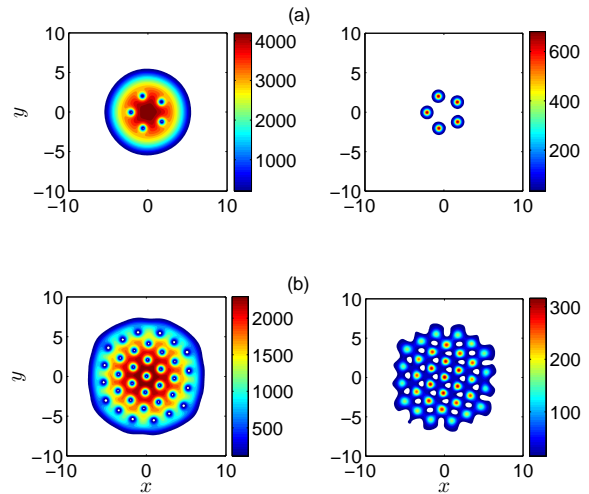}
\end{center}
\caption{(Color online) A series of density plots for component-1 (left column) and component-2 (right column). The parameters are $g_1=0.003$ and $g_2=0.006$ with $\eta=\xi=1$ and $\Gamma_{12}=0.3$ (which gives $g_{12}=0.0035$) and normalisation taken over the total density (Eq. (\ref{norm12})). The angular velocity of rotation is $\Omega$ and it takes the values (a) $0.5$ and (b) $0.9$. At these parameters the components are in region 2 of the phase diagram Fig. \ref{sketchv}. Distance is measured in units of $r_0$ and density in
units of $r_0^{-2}$.}
\label{3posv}
\end{figure}

{\it Region 2}. In the second region, only component-1 exists except for isolated density peaks that exist in component-2. These isolated density peaks occur at the same location as the vortices do in component-1, and are thus identical to the isolated coreless vortices described in detail in Sect. IV. Figure \ref{3posv} shows this case $\Gamma_{12}=0.3$ with $\Omega=0.5$ and $\Omega=0.9$.

{\it Region 3}. In the third region, only component-1 exists (the density peaks in component-2 that were present in region 2 are no longer present). Furthermore, only triangular vortex lattices are observed in this effective one component condensate (in component-1).
%Additionally, there are no signatures of these density peaks in a phase plot.

%{\bf Thomas Fermi approximation.}
Computations similar to Section III hold, except that now we have to
 set $\lambda=0$. This leads to $a_1/a_3=-\bar c_1/(2\bar c_2)$. We  see that this ratio (which
 is $S_z(0)$), reaches 1 or -1 when $\bar c_1=\pm2\bar c_2$. In our numerical case, this leads to $\bar \Gamma_{12}=0.5$.
 We see clearly 3 regimes: $\Gamma_{12}>\bar\Gamma_{12}$, where the condensates
are 2 disks, $\Gamma_{12}<0$, which is phase separation, in which case $S_z=1$ is the preferred state and
 $0<\Gamma_{12}<\bar \Gamma_{12}$, in which case the TF approximation leads to a computation of $\rho_T$ with
 coreless vortex lattices and variations in $S_z$ which improve the energy and lead to this intermediate state, still to be studied in more detail.

\section*{V. Conclusion}

We have presented phase diagrams of rotating two component condensates in terms of
 the angular velocity $\Omega$ and a nondimensionnalized parameter related to
 the coupling strengths
$\Gamma_{12}=1-g_{12}^2/(g_1g_2)$. We have analyzed
the various ground states and topological defects and have found
 four sets characterized by the
 symmetry preserving/symmetry breaking,
 coexistence or spatial separation of the components.
 When the geometry of the ground
states is either two disks (coexistence of components, region 1) or a disk
and an annulus (spatial separation keeping some symmetry, region 3), the topological
defects are coreless vortex lattices (with possible stabilization
of the square lattice) or giant skyrmions at the boundary
interface between the two components. In the complete symmetry breaking case, we have found vortex
 sheets and droplets. The difference of masses or coupling strengths between
 the components can induce very different patterns.

We have introduced an energy (\ref{sig}) related to the total density and a pseudo spin vector.
 The minimization in a generalized Thomas Fermi approximation provides a lot of information
 on the ground states for general masses, trapping frequencies and coupling strengths.
 Some parts of the phase diagrams can be justified rigorously, both in the case
  $\Gamma_{12}>0$, which had been studied before, but also in the case $\Gamma_{12}<0$
  with generalized models. This formulation of the energy
should bring in the future more information on the defects.

\section*{Acknowledgments}

The authors wish to thank Thierry Jolicoeur  for useful discussions
that took place for the duration of this work. They are very grateful to the referee for  his careful reading of the manuscript and his appropriate comments. We acknowledge
support from the French ministry Grant ANR-BLAN-0238, VoLQuan.


\begin{thebibliography}{1}
\bibitem{note2} For the parameters reported here, any vortices nucleated in the annulus of component-2 do not have corresponding density peaks outside of the circular component-1. However it is possible, using a different set of parameters, to create such density peaks outside of component-1.
\end{thebibliography}
\end{document}